\documentclass[
 aip,
 amsmath,amssymb,
 reprint,
 citeautoscript,
]{revtex4-1}

\usepackage{graphicx}
\usepackage{dcolumn}
\usepackage{bm}
\usepackage{braket}
\usepackage{mathtools}
\usepackage{xcolor}
\usepackage{mathrsfs}

\begin{document}

\title{Exploiting machine learning to efficiently predict multidimensional optical spectra in complex environments}

\author{Michael S. Chen}
\affiliation{Department of Chemistry, Stanford University, Stanford, California 94305, USA\looseness=-1}

\author{Tim J. Zuehlsdorff}
 \affiliation{Chemistry and Chemical Biology, University of California Merced, Merced, California 95343, USA\looseness=-1}

\author{Tobias Morawietz}
\affiliation{Department of Chemistry, Stanford University, Stanford, California 94305, USA\looseness=-1}

\author{Christine M. Isborn}
\email{cisborn@ucmerced.edu}
\affiliation{Chemistry and Chemical Biology, University of California Merced, Merced, California 95343, USA\looseness=-1}

\author{Thomas E. Markland}
\email{tmarkland@stanford.edu}
\affiliation{Department of Chemistry, Stanford University, Stanford, California 94305, USA\looseness=-1}

\date{\today}

\begin{abstract}
The excited state dynamics of chromophores in complex environments determine a range of vital biological and energy capture processes. Time-resolved, multidimensional optical spectroscopies provide a key tool to investigate these processes. Although theory has the potential to decode these spectra in terms of the electronic and atomistic dynamics, the need for large numbers of excited state electronic structure calculations severely limits first principles predictions of multidimensional optical spectra for chromophores in the condensed phase. Here, we leverage the locality of chromophore excitations to develop machine learning models to predict the excited state energy gap of chromophores in complex environments for efficiently constructing linear and multidimensional optical spectra. By analyzing the performance of these models, which span a hierarchy of physical approximations, across a range of chromophore-environment interaction strengths, we provide strategies for the construction of ML models that greatly accelerate the calculation of multidimensional optical spectra from first principles.
\end{abstract}

\maketitle

Chromophores and their photo-dynamics play a fundamental role in controlling biological functions, ranging from photosynthesis to visual perception, and in converting solar energy into the chemical energy stored in liquid solar fuels.\cite{Scholes2011, Schlau-Cohen2015, Luk2015, Ashford2015, Brennaman2016, Gulati2017, Yadav2018, Son2020} These essential processes are finely tuned by the interactions between a chromophore and its complex environment. Linear and multidimensional optical spectroscopies\cite{Mukamel00} probe the electronic transitions underlying these processes, providing insights into how a chromophore's environment tunes its energetics and optical response\cite{Engel2007,Calhoun2009,Dostal2016,Dean2016,Duan2017,Maiuri2018,Bolzonello2018}. Theoretical models can provide a direct link between these spectroscopic observables and the underlying electronic and atomic motions\cite{Adolphs2006,Muh2007, Olbrich2011,Shim2012,Lee2016,Lee2017,Segatta2017, Mallus2018, Blau2018, Jang2018,Loco_2018,Schnedermann2019,Segatta2019}. We have recently shown that methods based on a truncated cumulant expansion of the energy gap fluctuations provide a highly appealing approach to simulate optical spectra since they accurately capture vibronic and environmental effects for both strong and weak solvent interactions\cite{Zuehlsdorff2019}. However, the promise of using such advanced dynamics-based approaches to study linear and multidimensional optical spectroscopy is currently limited by the prohibitive cost of computing the lengthy time-sequence of electronic excitation energies required to converge the time correlation functions.

Machine learning (ML) offers the opportunity to dramatically reduce the cost of computing spectra by creating an efficient map between the chemical structure and the relevant spectroscopic properties\cite{Montavon2013,Ramakrishnan2015,Hase2016, Gastegger2017, Chen2018,Grisafi2018, Nebgen2018,Paruzzo2018,Pronobis2018,Sifain2018,Rodriguez2019,Christensen2019,Ghosh2019,Kananenka2019,Liu2019,Raimbault2019,Simine2019,Wilkins2019,Ye2019,Zhang2019,Li2020,Lu2020,Guan2020,Sommers2020,Kwac2020}. Here we develop and analyze the performance of three ML models for predicting electronic excitation energies of chromophores in solution, each model differing in how the environment is incorporated. By focusing on the anionic photoactive yellow protein chromophore (deprotonated $trans$-thiophenyl-$p$-coumarate, $\text{pCT}^-$) in water and the Nile red chromophore in both water and benzene, we demonstrate that our ML models trained on $\sim$2000 excited state electronic structure calculations can accurately capture linear and multidimensional optical spectra that would otherwise require orders of magnitude more computational effort. These ML frameworks enable the investigation of time-resolved spectroscopic simulations for a wide range of chromophore systems, creating new opportunities to connect experimental observables with electronic and atomistic dynamics.

\begin{figure*}[t!]
	\begin{center}
		\includegraphics[width=0.65\textwidth]{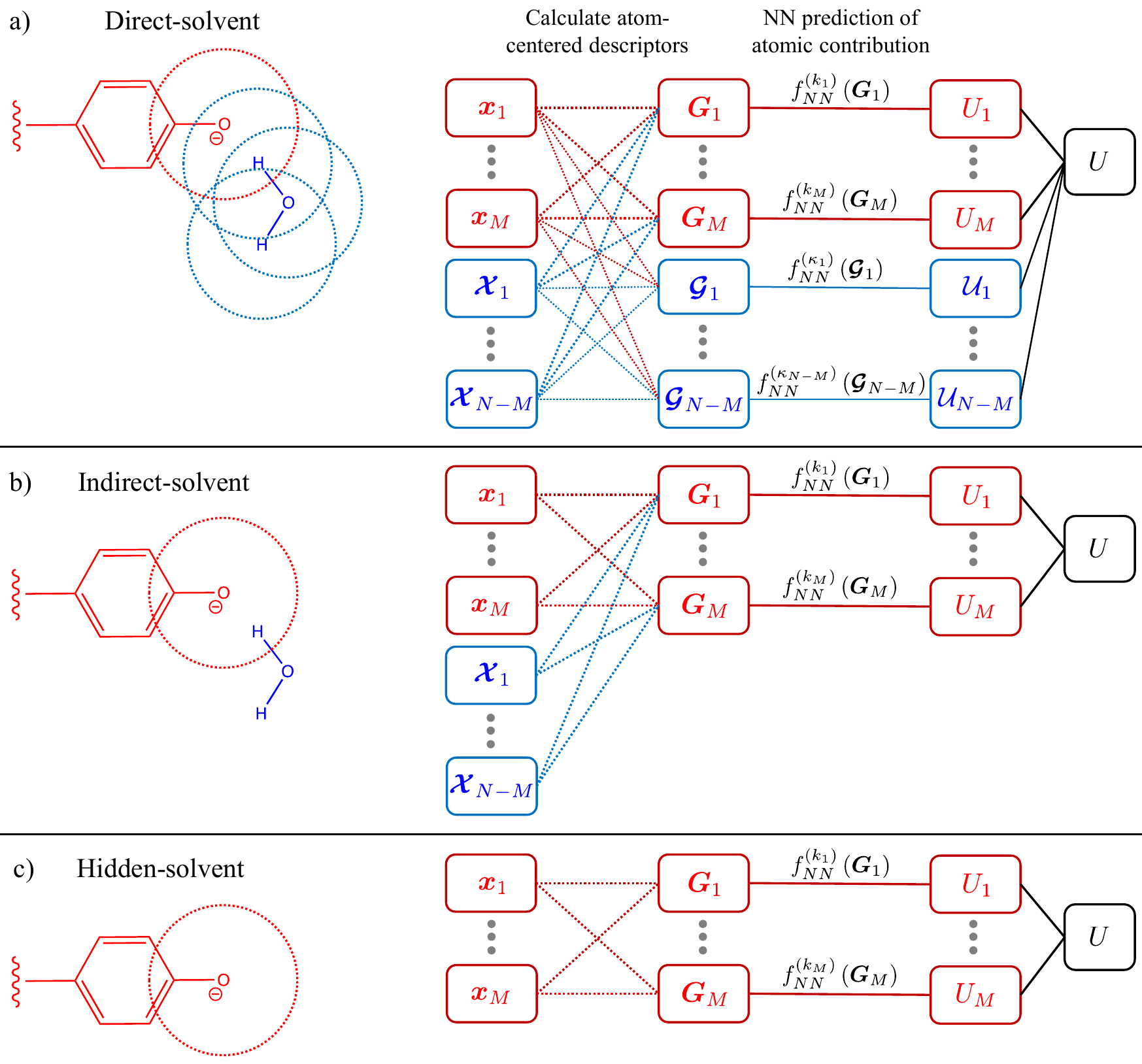}
	\end{center}
	\caption{Schematic depictions of the (a) direct-solvent, (b) indirect-solvent, and (c) hidden-solvent approaches to incorporating a chromophore's environment in the ML model.   The dotted circles represent the cutoff distance for the local environment encoded by a given atom-centered descriptor $\bm{G}_i$. For visualization purposes, these depicted cutoffs are much shorter than the 5~\AA\ cutoffs we actually employed. The connected rectangles illustrate the computational flow from chromophore and environment atomic positions and types $\{\bm{x}, \bm{\mathcal{X}} \}$ to excited state energy gap prediction $U$ for the three ML models. The neural networks for a particular element type, $k_i$, are denoted by $f_{NN}^{(k_i)}$ and they predict an atomic energy contribution to the energy gap $U_i$ given an atom-centered descriptor $\bm{G}_i$ for atom $i$.}  
	\label{fig:model-diagrams}
\end{figure*}

Atom-centered ML approaches \cite{Behler2007,Bartok2010,Artrith2011,Behler2011,Bartok2013,Morawietz2016,Smith2017,Artrith2017,Schutt2017,Schutt2017a,Yao2018,Zhang2018,Grisafi2018,Gastegger2018,Lubbers2018,Willatt2019,Unke2019,Artrith2019} have been used to generate spectra, such as Raman and infra-red, via evolution on a ML ground state potential energy surface \cite{Morawietz2018,Morawietz2019,Sommers2020,Gastegger2017,Nebgen2018,Sifain2018,Zhang2019} and have also been used to predict spectroscopic properties directly \cite{Gastegger2017,Chen2018,Grisafi2018,Nebgen2018,Paruzzo2018,Pronobis2018,Sifain2018,Christensen2019,Ghosh2019,Kananenka2019,Liu2019,Raimbault2019,Wilkins2019,Zhang2019,Lu2020,Sommers2020,Kwac2020}. Here we introduce an atom-centered ML approach to model the electronic energy gap $U$, the difference in energy between the ground and first excited state. As is the case for ML potentials that predict the potential energy of a system, the energy gap is given as a sum of atomic contributions, $U_i$, obtained from a neural network specific to the $i^{\text{th}}$ atom's element, $f_{NN}^{(k_i)}$, where $k_i$ is its element type. The input to an element specific neural network is a set of descriptors $\bm{G}_i$ that represent the local chemical environment around atom $i$. We employ atom-centered Chebyshev polynomial descriptors to encode the positions of nuclei around a given atom\cite{Artrith2017} in a way that incorporates translational and rotational symmetries while also being systematically improvable. Within this general ML framework we compare three different ways of including the environment and analyze the physical effects each model can and cannot capture. Figure~\ref{fig:model-diagrams} summarizes these three approaches.

The direct-solvent approach, Figure~\ref{fig:model-diagrams}a, represents the most straightforward and ``brute-force" application of an ML framework. In this approach, both chromophore and solvent atoms are treated equivalently in the ML model and an atomic contribution to the energy gap is calculated for each atom in the system. The total energy gap, $U$, for a system of $N$ atoms is the sum of contributions from $M$ chromophore and $N-M$ solvent atoms,
\begin{equation} \label{eqn:direct-solvent}
\begin{split}
    U = & \sum_{i=1}^M U_i + \sum_{j=1}^{N-M} \mathcal{U}_j \\
    = & \sum_{i=1}^M f_{NN}^{(k_i)} \left( \bm{G}_{i} ( \{ \bm{x} , \bm{\mathcal{X}} \} ) \right) +\sum_{j=1}^{N-M} f_{NN}^{(\kappa_j)} \left(\bm{\mathcal{G}}_{j}(\{\bm{x}, \bm{\mathcal{X}}\}) \right),
\end{split}
\end{equation}
\noindent where $U_i$ and $\mathcal{U}_j$ are contributions from the $i^{\text{th}}$ chromophore and $j^{\text{th}}$ solvent atoms, respectively. The position $\bm{r}_i$ and element type $k_i$ of a given chromophore atom $i$ are denoted as $\bm{x}_i \equiv (\bm{r}_i, k_i)$ and similarly for solvent atoms, $\bm{\mathcal{X}}_j \equiv (\bm{\mathcal{R}}_j, \kappa_j)$. The full set of all chromophore and solvent atom positions and element types is thus denoted $\{\bm{x}, \bm{\mathcal{X}}\}$. These serve as inputs for calculating the respective chromophore, $\bm{G}_i$, and solvent, $\bm{\mathcal{G}}_j$, atom-centered descriptors.

Because chromophore excitations are often localized, not all atoms in these systems will contribute equally to the energy gap. For example, in the chromophores explored here, the changes to the electron density upon excitation are mostly localized within the $\pi$ system\cite{Zuehlsdorff2020}. Hence, other atoms not involved in the $\pi$ conjugation, including solvent atoms, are not as important in determining the excitation energy. Given enough data and a sufficiently flexible set of neural networks, a direct-solvent ML model would learn to appropriately weight the different contributions. However, owing to the large computational cost of excited state electronic structure calculations, the ideal ML model should be trainable using a minimal number of energy gaps. By explicitly incorporating simple physical approximations into the ML model, we can offload some of the physics the ML model must learn and let it focus on capturing less intuitive structure-property relationships. Consequently, these simplified and more focused ML models can be more data efficient, i.e. less data will be needed to train an accurate model. Here we leverage the locality of electronic excitations to introduce an indirect-solvent approach (Fig.~\ref{fig:model-diagrams}b) where only the contributions $U_i$ from chromophore atoms are considered explicitly. The positions and element types of the solvent atoms $\{\bm{\mathcal{X}}\}$ still influence the energy gap, albeit indirectly through the chromophore atoms' local environment descriptors $\bm{G}_i$,
\begin{equation} \label{eqn:indirect-solvent}
    U = \sum_{i=1}^M U_i = \sum_{i=1}^M f_{NN}^{(k_i)}\left(\bm{G}_i(\{\bm{x}, \bm{\mathcal{X}}\})\right).
\end{equation}
\noindent The physical assumptions underlying the construction of this model are that solvent atoms located far from the chromophore (defined by the cutoff for the chromophore atom-centered descriptors $G_i$) and solvent-solvent interactions contribute negligibly to the excited state electronic energy gap. If these assumptions hold, this indirect-solvent model should be more data efficient than the direct-solvent model.

A more drastic simplification of the model would be to completely neglect solvent atom positions, as shown in the hidden-solvent model in Figure~\ref{fig:model-diagrams}c. Like in the indirect-solvent model, the total energy gap in this model is simply a sum of contributions from only the chromophore atoms (Eq.~\ref{eqn:indirect-solvent}). However, in the hidden-solvent model $\bm{G}_i$ is only a function of chromophore atom positions and types $\{\bm{x}\}$, i.e. $\bm{G}_i(\{\bm{x}, \bm{\mathcal{X}}\}) \to \bm{G}_i(\{\bm{x}\})$. By training on electronic excitation energies generated in the presence of solvent, this model can capture an average solvatochromic shift, but will be unable to connect any solvation effects on the energy gap to specific solvent configurations. Hence, this model will likely fail for systems where the chromophore interacts strongly in a site-specific manner with its environment e.g., via hydrogen bonding. On the other hand, due to its simplicity, this model should require less training data to saturate its accuracy than the other two models that encode atomistic information for the environment.

\begin{figure}[hbt!]
	\begin{center}
		\includegraphics[width=0.425\textwidth]{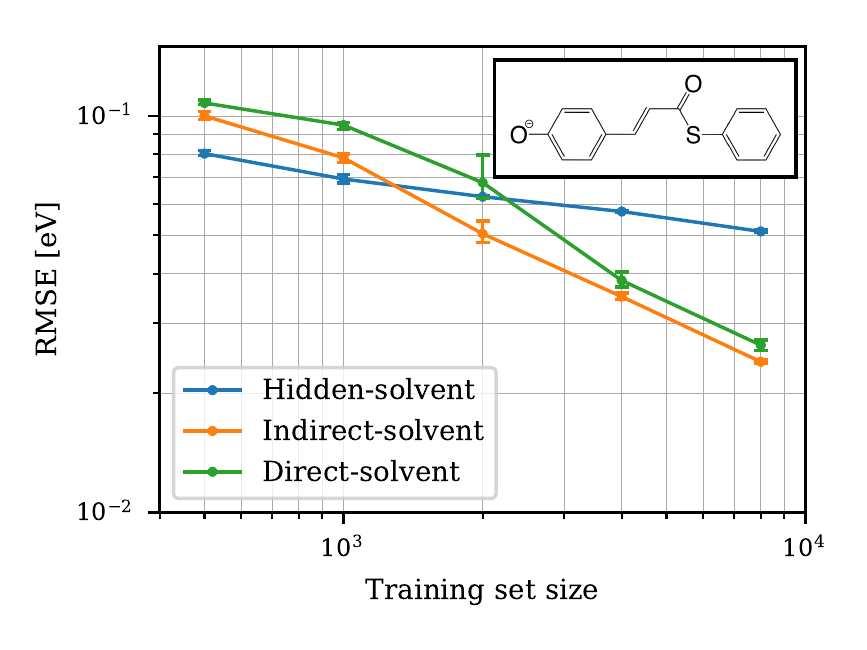}
	\end{center}
    \caption{Learning curves for $\text{pCT}^-$ in water show the crossover in the RMSE validation set error for the ML solvation models as the training set size is increased. For small training sets the hidden-solvent approach gives the lowest errors whereas at larger training set size the indirect-solvent method performs best. Points represent an average over four training realizations and the bars represent the range of validation errors, the lower bar representing the minimum error and the upper bar representing the maximum error.}
	\label{fig:pyp_water_learningcurves}
\end{figure} 

To assess these ML solvation models across a range of chromophore-solvent interaction strengths, we used the open source {\ae}net package\cite{Artrith2016} to train models for the $\text{pCT}^-$ chromophore in water (strong, site-specific interactions between the anionic chromophore and water), the Nile red chromophore in water (medium strength interactions), and the Nile Red chromophore in benzene (weak interactions) (SI Appendix~1). We analyze the data efficiency of the models by assessing their energy gap prediction errors as a function of training set size. We also assess how accurately the ML models reproduce the linear and multidimensional optical absorption spectra as compared to those obtained by using the energy gaps computed via the reference \textit{ab initio} electronic structure method (SI Appendix~1). We compute these spectra using a cumulant expansion approach truncated at second order (SI Appendix~2), \cite{Mukamel1995,Mukamel1985} which we have previously shown gives accurate optical spectra in both strong and weak solvent coupling regimes \cite{Zuehlsdorff2019}. In comparatively analyzing our results, we connect the ML solvation models' failures to their inability to capture how certain physical effects manifest in the optical spectra.

\begin{figure*}[hbt!]
	\begin{center}
        \includegraphics[width=0.85\textwidth]{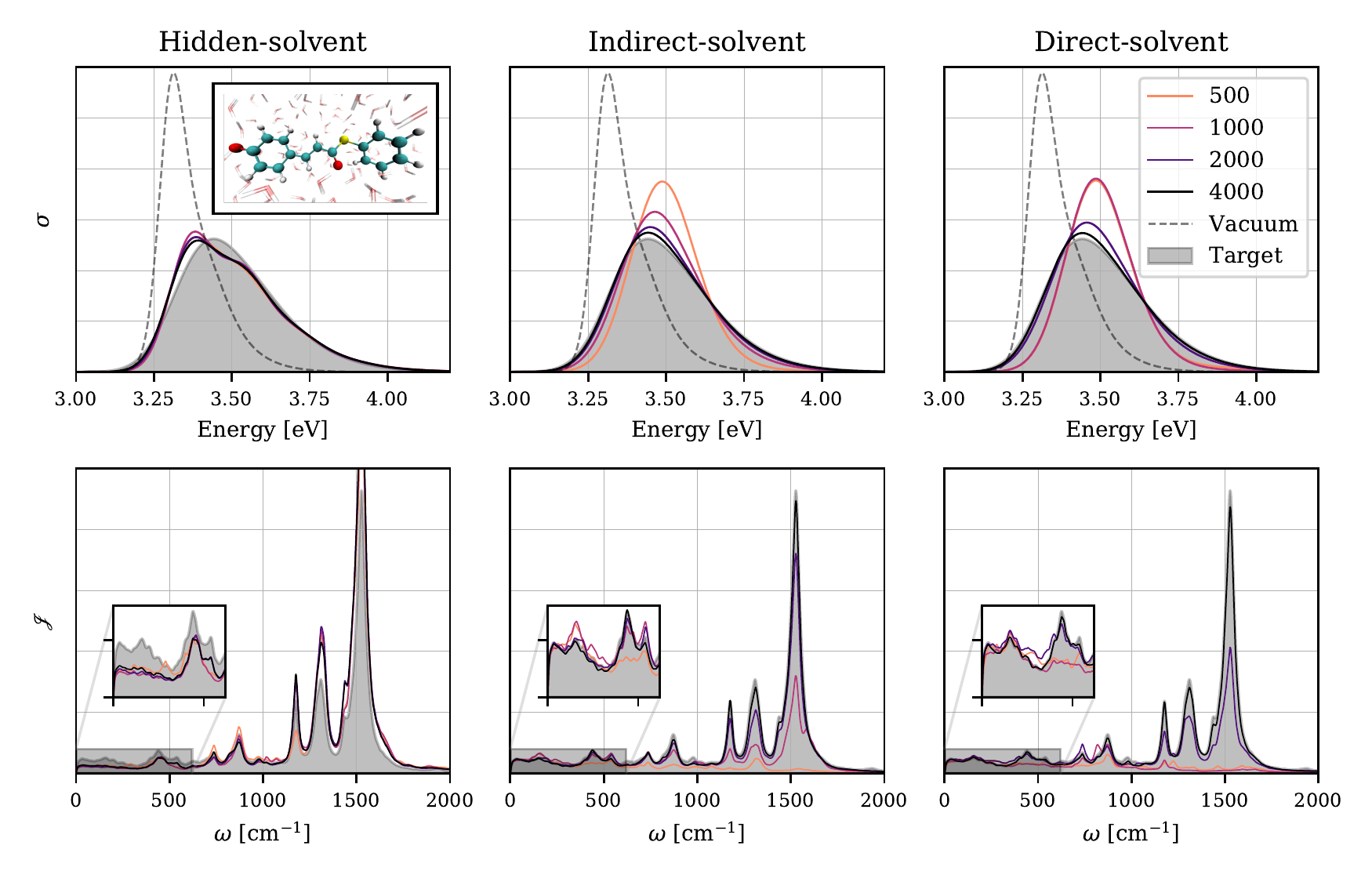}
	\end{center}
	\caption{Linear absorption spectra (top) and spectral densities (bottom) for the three ML solvation models showing the convergence with respect to training set size ($\text{pCT}^-$ in water). The indirect- and direct-solvent models achieve graphical accuracy with large numbers of training points, but the hidden-solvent model never achieves this level agreement. Shading represents the target result as computed using 32,000 electronic structure energy gaps. The trained ML models were used to predict an energy gap for the same 32,000 structures and those predictions were used to compute the spectral densities and linear spectra.}
	\label{fig:pyp_water_specdens_absspec}
\end{figure*}

The $\text{pCT}^-$ chromophore solvated in water is an example of an anionic chromophore that hydrogen bonds to the solvent. This serves as a particularly challenging test system since the interactions between the chromophore and solvent are sufficiently strong such that both implicit and molecular mechanics treatments of the environment fail to accurately capture the linear absorption spectrum \cite{Provorse2016,Isborn2012,Zuehlsdorff2016,Milanese2017,Zuehlsdorff2020}. Figure~\ref{fig:pyp_water_learningcurves} demonstrates the compromise between the accuracy and the amount of training data for each of the ML solvation models. As the training set size increases, the root mean squared errors (RMSE) over the validation set (SI Appendix~3) for the three ML models exhibit crossovers, with the hidden-solvent method yielding the lowest error at the smallest training set size (500 training points) and the indirect-solvent model becoming the most accurate by 2000 training points with a RMSE (0.048~eV) that is considerably smaller than the standard deviation of the reference energy gaps (0.123~eV, SI Appendix~6). Both direct and indirect-solvent models are more accurate than the hidden-solvent model if at least 3000 training points are used. These crossovers can be rationalized in terms of the canonical trade-off between accuracy and variance: more complex models are more accurate when provided enough data but are prone to overfitting when using small training sets, whereas simpler models are less susceptible to overfitting but cannot capture the full physics of the system.\cite{Geman1992} Here, the direct-solvent model has the greatest complexity and can in principle incorporate the most physics while the hidden-solvent model applies a drastic simplification. Our indirect-solvent model is a compromise between these two extremes, balancing model complexity with data efficiency. This balance is reflected in the learning curve where the indirect-solvent model outperforms the other two in the range of 2000-8000 training points. Since the direct-solvent model can be trained to learn the same approximations explicitly enforced in the indirect-solvent model, we expect that with enough data and a sufficiently flexible set of neural networks it should approach and eventually surpass the accuracy of the indirect-solvent model. However, in this case, even with 8000 training points the direct-solvent model is still marginally less accurate than the indirect-solvent model.

To provide insight into how the errors in the ML-predicted energy gaps manifest in the linear optical absorption spectrum, Figure~\ref{fig:pyp_water_specdens_absspec} shows the linear spectra (upper panels) for $\text{pCT}^{-}$ in water using training set sizes ranging from 500 to 4000 for each of the three ML models. Even when trained on the smallest dataset, all the ML models capture the spectrum significantly more accurately than if solvation effects had been completely neglected (i.e. the spectrum for the chromophore in vacuum). The direct-solvent approach converges systematically to the target result and obtains graphical agreement with the largest training set reported (4000), whereas the indirect-solvent approach converges to a similar level of accuracy with half as many training points. On the other hand, the hidden-solvent approach gives a consistent spectrum for all the training set sizes but exhibits a spurious shoulder. It is worth noting that the 500 training points needed to converge the optical spectrum for the hidden-solvent model, where only the chromophore atoms positions are fed into the network, is consistent with how many are required to converge a model for $\text{pCT}^-$ in vacuum (SI Appendix~5).

\begin{figure*}[hbt!]
	\begin{center}
        \includegraphics[width=0.85\textwidth]{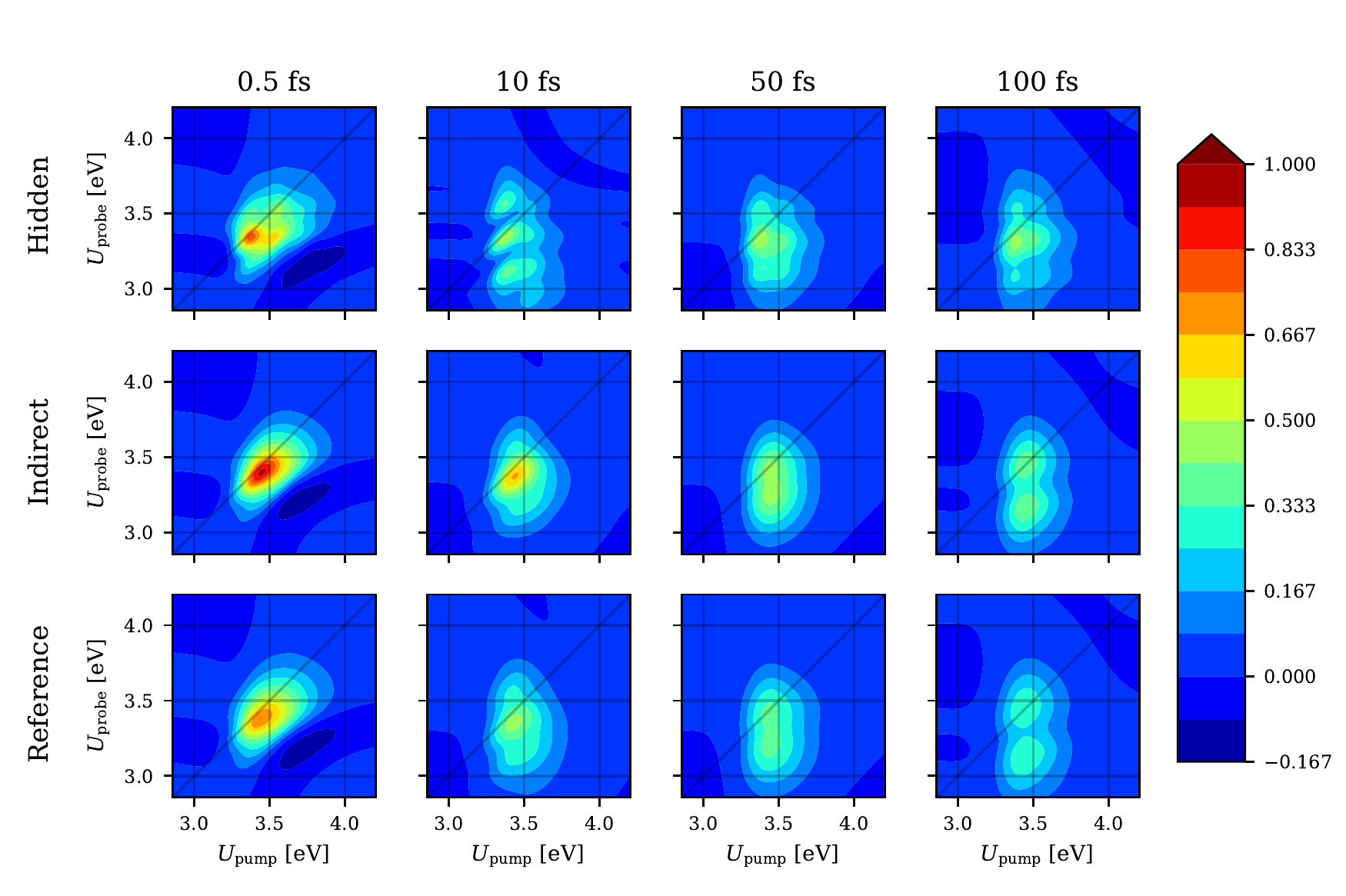}
	\end{center}
	\caption{Two-dimensional electronic spectra (2DES) of $\text{pCT}^-$ in water showing the ability (inability) of the indirect-solvent (hidden-solvent) model to accurately capture spectral diffusion. A series of time delays are reported ranging from 0.5-100~fs for both the hidden-solvent (top) and indirect-solvent (middle) models, which are trained on 2000 energy gaps then used to predict the 32,000 excitation energies used to compute these spectra. The corresponding reference results (bottom) are computed using 32,000 energy gaps from electronic structure calculations.}
	\label{fig:pyp_water_2DES}
\end{figure*}

The physical origin for these differences can be understood by examining the spectral densities in the lower panels of Figure~\ref{fig:pyp_water_specdens_absspec}, which are intimately linked to their absorption spectra via the line-shape function (SI Eq.~3). These spectral densities encode the coupling of the system's energy gap to its vibrational modes, yielding a frequency distribution of energy gap fluctuations. The higher frequency peaks correspond to the coupling of the energy gap to intrachromophore modes while the lower frequency peaks are a result of slower solvent-coupled modes\cite{Zuehlsdorff2020}. From Figure~\ref{fig:pyp_water_specdens_absspec}, we see that the hidden-solvent model, which neglects the solvent positions, systematically overpredicts the intensities of the high frequency intrachromophore modes while underestimating the lower frequency solvent modes (see inset in Figure~\ref{fig:pyp_water_specdens_absspec}). These errors manifest in the linear absorption spectrum as an over-accentuated vibronic shoulder. The ML models that explicitly include solvent positions are able to capture the intensities of both the solvent-coupled and intrachromophore modes with the indirect-solvent model needing as few as 2000 training points. Below that number, the indirect-solvent model underestimates the intensity of the high frequency modes in the spectral density, resulting in absorption spectra that are missing the high-energy vibronic tail and that are overly symmetric. For the less data efficient direct-solvent model, these same errors become apparent below 4000 training points. Despite the large discrepancies in the high frequency region of the spectral density, these errors cause small changes in the linear absorption spectrum due to the $1/\omega^{2}$ factor in SI Eq.~3. However, as we demonstrate below, these same errors manifest more prominently in the corresponding multidimensional spectra.

The photo-dynamics of chromophores are often characterized using multidimensional spectroscopy. Therefore it is important our models accurately reproduce these spectra, such as those generated from two dimensional electronic spectroscopy (2DES)\cite{Mukamel00,Jonas2003,Cho2008}. 2DES provides a more stringent and information-rich test than linear spectroscopy; at short time delays, spectra show broadened vibronic features attributable to fast intrachromophore vibrational modes and, at longer time delays, they probe the time scales of slower solvent-coupled relaxation processes. The 2DES reference spectra in Figure~\ref{fig:pyp_water_2DES} show that after a short time delay of 0.5~fs the peak is elliptical and fairly symmetric along the diagonal since the system has not had much time to evolve. By 10~fs, the 2DES exhibits off-diagonal vibronic peaks attributable to the high-frequency vibronic modes of the chromophore. At the longest time delay (100~fs), there are two distinct peaks with the diagonal peak corresponding to ground-state bleaching and the peak below the diagonal corresponding to stimulated emission (see SI Appendix~10). Intrachromophore and solvent reorganization stabilize the excited-state and thus induce a Stokes shift in the stimulated emission\cite{Lee2017a,Hybl2001a,Hybl2002}. For this system, the calculated reference reorganization energy is $\lambda = 0.284~\text{eV}$ (see SI Appendix~4) and at long time delays the stimulated emission peak will be Stokes shifted $\approx2\lambda$ below the diagonal\cite{Mukamel1995,Sun2016}.

Turning to the ML models, Figure~\ref{fig:pyp_water_2DES} shows the more stringent test provided by 2DES makes the failures of the hidden-solvent model apparent; it gives a qualitatively incorrect single spectral feature at 100~fs. This failure is consistent with the hidden-solvent model's spectral density (Fig.~\ref{fig:pyp_water_specdens_absspec}), where the intensity of the low frequency features corresponding to solvent motion are underestimated. In the corresponding 2DES, this error manifests in the enhanced vibronic structure at 10~fs and in an underestimation of the Stokes shift at 100~fs, as evinced by the failure to separate the ground state bleach and stimulated emission signals. The lack of intensity in the low frequency part of the spectral density means that less vibrational energy can be transferred from fast intrachromophore modes to collective environment modes, thus leading to a reduced Stokes shift in the limit of long delay times.  

In contrast, the indirect-solvent model trained on only 2000 data points accurately captures the shapes and relative intensities of the reference 2DES. At a time delay of 100~fs, the peak separation for the indirect-solvent 2DES is 0.28~eV, which is in good agreement with the reference peak separation of 0.31~eV. In part, this good agreement reflects the accurate reproduction of the reference reorganization energy by the indirect-solvent model ($\lambda = 0.262~\text{eV}$ vs. $\lambda = 0.284~\text{eV}$ for the reference). When smaller training sets are used (SI Fig.~6), the accuracy of the indirect-solvent 2DES degrades. The most obvious failure is the absence of the vibronic cross-peaks at a 10~fs time delay, which is consistent with the underestimation of the high frequency intrachromophore modes in the spectral density.

\begin{figure}[hbt!]
	\begin{center}
		\includegraphics[width=0.425\textwidth]{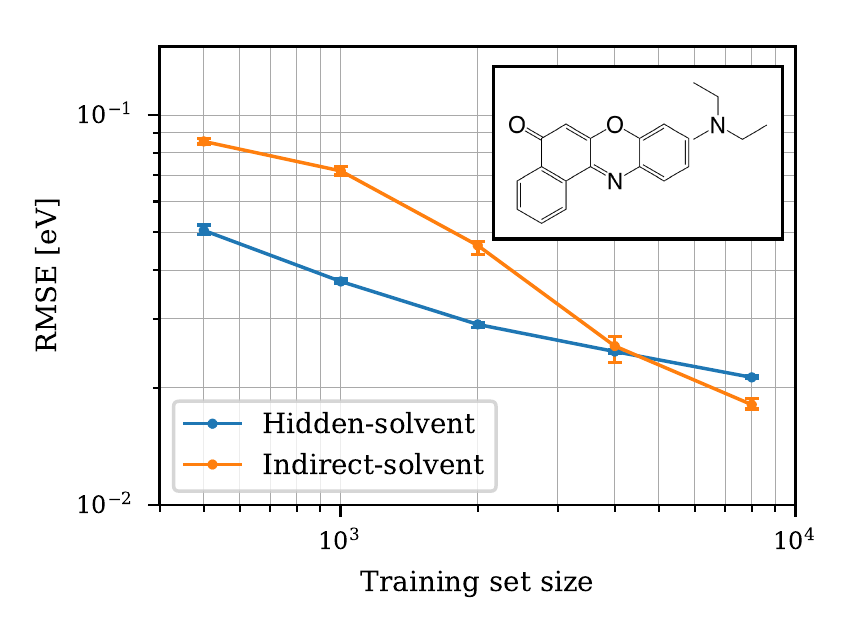}
	\end{center}
	\caption{Learning curves for Nile red in benzene showing how the hidden-solvent model outperforms the indirect-solvent model up until 4000 training points.}
	\label{fig:nilered_benzene_learningcurves}
\end{figure}

\begin{figure}[hbt!]
	\begin{center}
		\includegraphics[width=0.5\textwidth]{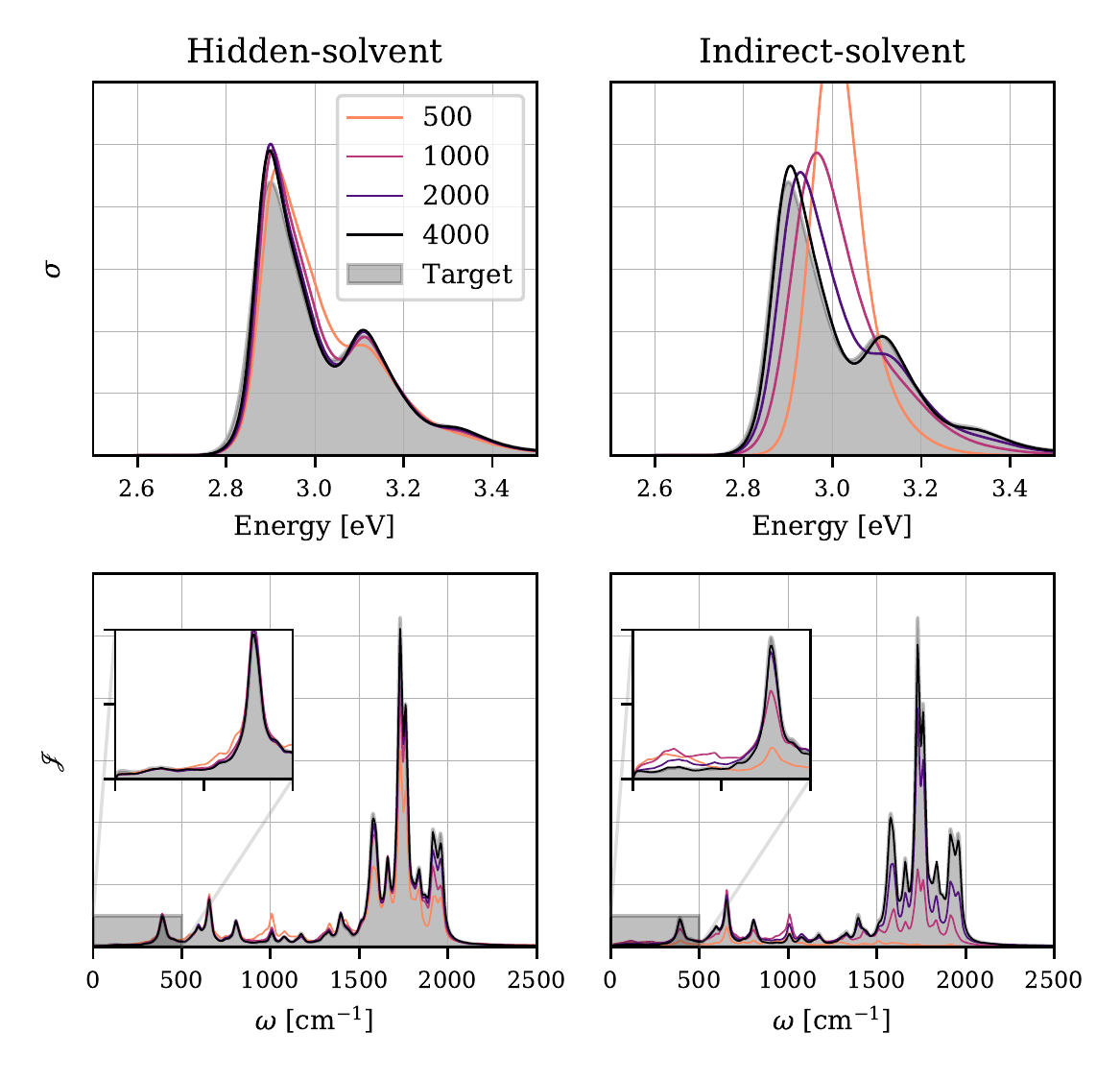}
	\end{center}
	\caption{Linear absorption spectra (top row) and spectral densities (bottom row) for Nile red in benzene as a function of training set size using the hidden- and indirect-solvent models. Shading represents the target as computed using 30,000 electronic structure excitation energies.}
	\label{fig:nilered_benzene_specdens_absspec}
\end{figure}

\begin{figure*}[hbt!]
	\begin{center}
		\includegraphics[width=0.85\textwidth]{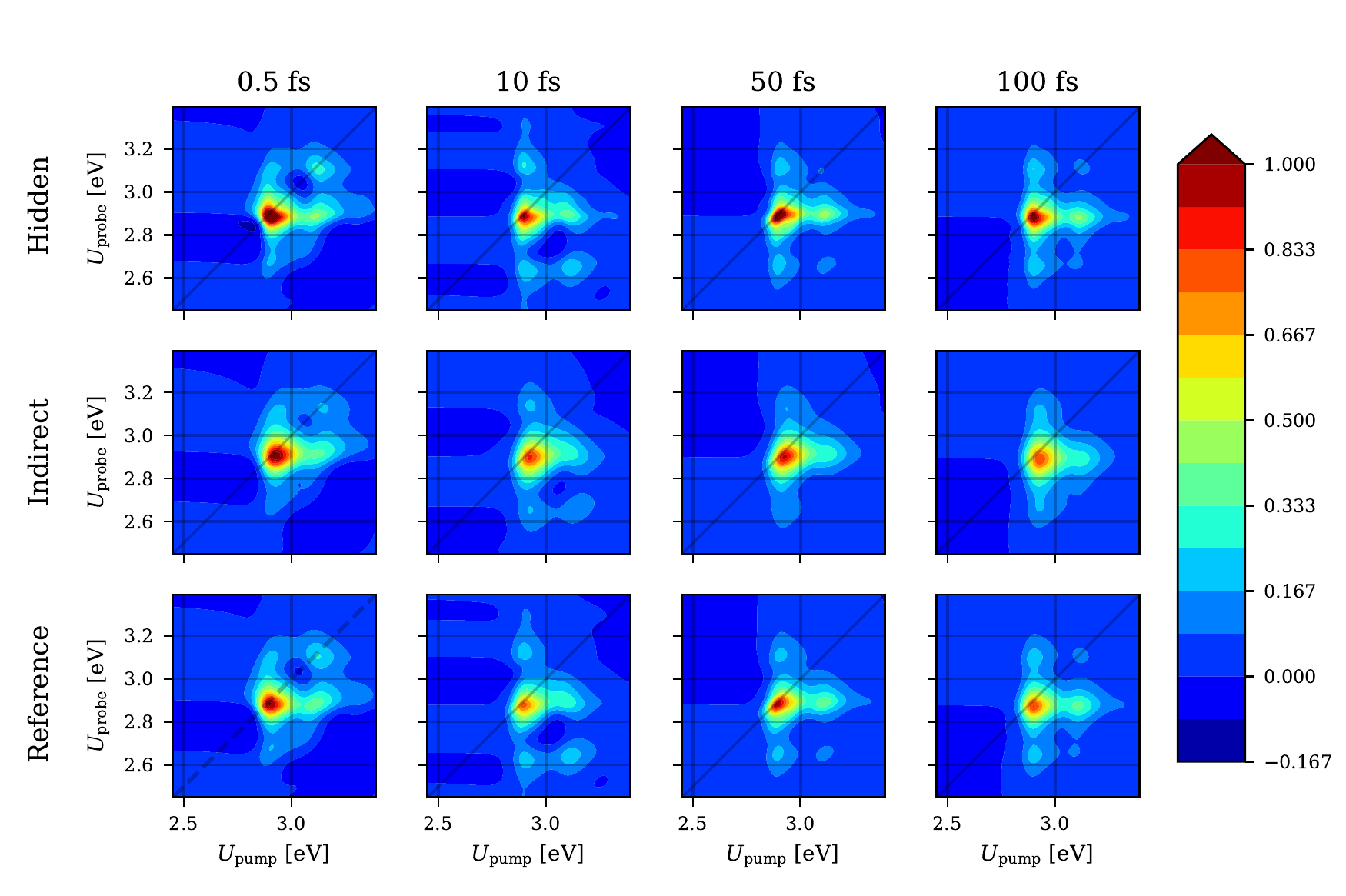}
	\end{center}
	\caption{2DES for Nile red in benzene at time delays of 0.5, 10, 50, and 100~fs using the hidden- and indirect-solvent models as trained on 2000 points. The reference 2DES was computed using 30,000 electronic structure excitation energies.}
	\label{fig:nilered_benzene_2DES}
\end{figure*}

To address the transferability of these ML frameworks, we now consider the Nile red chromophore solvated in benzene and water representing a weakly and more strongly interacting chromophore-solvent system, respectively. Given that the indirect-solvent approach outperformed the direct method for all training set sizes in the case of $\text{pCT}^-$ in water, we now focus primarily on the indirect-solvent approach and how its results compare with those of the hidden-solvent approach, particularly in the case of weaker chromophore-solvent coupling.

The learning curves for Nile red in benzene in Figure~\ref{fig:nilered_benzene_learningcurves} demonstrate that the hidden-solvent approach outperforms the indirect-solvent approach for a larger range of training set sizes when the chromophore and solvent are more weakly coupled. The RMSE of the hidden-solvent model for Nile red in benzene using 500 training points is lower than that obtained using 8000 points for $\text{pCT}^-$ in water and gives good agreement with the linear optical spectrum (Fig.~\ref{fig:nilered_benzene_specdens_absspec}). Because of the strong performance of the hidden-solvent approach, it takes 4000 training points before the indirect-solvent approach can match its accuracy. Similarly, the 2DES demonstrates that the predictions from the hidden-solvent model trained on only 2000 points reproduce the time evolution of the vibronic spectral features (Fig.~\ref{fig:nilered_benzene_2DES}). In contrast, the indirect-solvent model requires 4000 training points to generate 2DES of similar accuracies (SI Fig.~7). These results indicate that the interaction between Nile red and benzene is sufficiently weak such that explicit incorporation of solvent atoms in the ML model is not necessary and actually hampers training by introducing unnecessary complexity.

For a system of intermediate chromophore-environment interaction strength, we consider Nile red in water. As shown in SI Figures~8 and 9, the hidden-solvent model yields optical spectra for Nile red in water that over-accentuate vibronic features regardless of how many training points are used, consistent with the results for strongly interacting $\text{pCT}^-$ in water. The indirect-solvent model trained on 4000 points qualitatively reproduces the 2DES spectra for Nile red in water, but for smaller numbers of training points the spectrum exhibits a loss in off-diagonal vibronic features as is most evident at the 10~fs delay time. This phenomenon results from the spectral densities where the intensities of the high-frequency intrachromophore modes are underestimated when the indirect-solvent model is trained on small datasets. When trained on 4000 points, this error in the spectral density is rectified and consequently the corresponding indirect-solvent 2DES is in good agreement with the reference. Taken all together, our results across these different systems show that it is the chromophore-environment interaction strength that determines whether the indirect-solvent model outperforms the hidden-solvent model.

Here we have demonstrated that accurate linear and multidimensional optical spectra for solvated chromophores can be efficiently calculated using ML models to predict the excited-state electronic energy gaps. We have introduced and analyzed how different solvent representations in ML models can be more data efficient and accurate depending on the strength of the chromophore-solvent coupling. For weakly coupled systems like Nile red in benzene, we were able to completely ignore solvent positions in the ML model and trained a model which gave accurate linear and 2D electronic spectra with as few as 2000 training points. For systems with stronger and more site specific chromophore-solvent interactions, like $\text{pCT}^-$ in water, the simplified hidden-solvent approach did not suffice. However, by accounting for solvent atoms located close to the $\text{pCT}^-$ chromophore, we were able to train an accurate indirect-solvent model with 2000 training points. This work therefore provides a highly efficient scheme to generate accurate multidimensional optical spectra for chromophores in condensed phase environments. For example, the 2DES of the $\text{pCT}^-$ chromophore with a surrounding solvent shell of 166 water molecules (527 atoms) required 32,000 energy gaps to compute. Each one of those 32,000 reference excited-state electronic structure calculations takes $~$8 hours on a Nvidia Tesla K80 GPU. In contrast, our ML models can predict the energy gaps for all 32,000 configurations in one minute using a 32 core CPU node. Hence, the reduction from 32,000 to 2000 calculations represents a dramatic computational savings of 240,000 GPU hours. These massive savings can be leveraged to reduce the errors in the optical spectra that arise from other sources, such as incomplete statistical sampling of the ground state surface\cite{Zwier2007,Loco2019}, the quantum dynamics approach, or in the level of electronic structure used to compute the energy gaps\cite{Isborn2013}, all of which can introduce larger errors than those in our ML model trained with only 500 points.

Finally, although we focused solely on predicting excited state energy gaps for solvated chromophores, the ML frameworks we present are generally applicable to any scalar structure-based property where a natural separation exists between a molecule and its surroundings. We also note that the labeling of the atoms belonging to the molecular center of interest and the atoms belonging to the environment is flexible. For instance, if one has pre-existing knowledge of where the electronic excitation is localized on the molecule (e.g. only the $\pi$ system) then the data efficiency of the indirect-solvent model could be improved by leveraging that intuition and further localizing the model to those specific parts of the chromophore.  Similarly, if certain atoms in the environment couple strongly to the electronic excitation, they can be included with the atoms in the ML model that directly contribute to the excitation energy.

In conclusion, we have shown that ML models can accurately predict excited state properties of chromophores in complex environments, leading to dramatically reduced computational cost for simulating nonlinear optical spectra. This ability will enable the combination of ML with advanced semiclassical and quantum dynamical methods to study the photo-dynamics and multidimensional spectroscopy of chromophores in complex environments. This work therefore provides a physically informed and data efficient ML-based route to alleviate the computational bottleneck in the calculation of linear and multidimensional optical spectroscopies for a wide range of chemical systems.

\acknowledgments
We greatly thank Harish Bhat, Andr\'es Montoya-Castillo, and Yuezhi Mao for useful conversations. We also thank Nongnuch Artrith for providing us access to the developer version of {\ae}net. This work was supported by the U.S. Department of Energy, Office of Science, Office of Basic Energy Sciences under Award Number DE-SC0020203. T.E.M also acknowledges support from the Camille Dreyfus Teacher-Scholar Awards Program. This work used the XStream computational resource, supported by the National Science Foundation Major Research Instrumentation program (ACI-1429830).

\section*{Author Contributions}
All authors contributed to the design of this research. M.S.C, T.J.Z., and T.M. performed the research. All authors were involved in the analysis of results and writing of the manuscript.

\end{document}


\title{Supplementary Information: Exploiting machine learning to efficiently predict multidimensional optical spectra in complex environments}

\author{Michael S. Chen}
\affiliation{Department of Chemistry, Stanford University, Stanford, California 94305, USA\looseness=-1}

\author{Tim J. Zuehlsdorff}
 \affiliation{Chemistry and Chemical Biology, University of California Merced, Merced, California 95343, USA\looseness=-1}

\author{Tobias Morawietz}
\affiliation{Department of Chemistry, Stanford University, Stanford, California 94305, USA\looseness=-1}

\author{Christine M. Isborn}
\email{cisborn@ucmerced.edu}
\affiliation{Chemistry and Chemical Biology, University of California Merced, Merced, California 95343, USA\looseness=-1}

\author{Thomas E. Markland}
\email{tmarkland@stanford.edu}
\affiliation{Department of Chemistry, Stanford University, Stanford, California 94305, USA\looseness=-1}

\date{\today}

\keywords{}

\maketitle

\tableofcontents
\newpage

\section{Simulation Details} 
\label{app:simulations}
For pCT$^-$ in vacuum, \textit{ab initio} molecular dynamics trajectories were taken from the data set generated in Ref.~\onlinecite{Zuehlsdorff2018}. The original data set consists of four independent trajectories of 25~ps length. For the present study, we extracted snapshots every 2~fs from the last 16~ps of each trajectory, yielding a total of 36,000 snapshots. For each of those snapshots, vertical excitation energies were computed using time-dependent density functional theory (TDDFT) in the Tamm-Dancoff approximation\cite{Hirata1999} as implemented in the TeraChem code\cite{Isborn2011}. Calculations were carried out with the CAM-B3LYP functional\cite{Woon1995,Yanai2004} and the 6-31+G* basis set.

For pCT$^-$ in water, \textit{ab initio} molecular dynamics trajectories of the chromophore in a box of 166 water molecules were taken from the data set generated in a previous study\cite{Zuehlsdorff2018}. In total, 8~ps trajectories were extracted from 8 of the 10 independent 15~ps trajectories. From these 8 independent trajectories, snapshots were extracted every 2~fs, yielding 4000 snapshots per trajectory and 32,000 snapshots in total. For each of these snapshots, the vertical excitation energy was computed using time-dependent density functional theory (TDDFT) in the Tamm-Dancoff approximation\cite{Hirata1999} as implemented in the TeraChem code\cite{Isborn2011}. Calculations were carried out with the CAM-B3LYP functional\cite{Woon1995,Yanai2004} and the 6-31+G* basis set.  All 166 water molecules were treated fully quantum mechanically (QM) in the calculation, corresponding to 527 atoms in the QM region. To ensure that the water molecules at the edge of the simulation box are embedded in a realistic electrostatic environment, the entire explicit solvent box was embedded in a polarizable continuum model (PCM)\cite{Tomasi2005} with a static dielectric constant of $\epsilon_0=80.0$. 

For Nile red in benzene and in water, force-field based molecular dynamics simulations were carried out using AMBER. The force field parameters, solvent box size, simulation parameters and details of the equilibration were identical to those reported in a recent study\cite{Zuehlsdorff2020}. Three independent trajectories, each separated by 1~ns, of 20~ps in length were generated and snapshots were extracted every 2~fs, yielding a total of 30,000 data points for each system. TDDFT calculations were carried out with the CAM-B3LYP functional and the 6-31+G* basis set, a basis set that has been shown in a previous study to be sufficiently large to yield well-converged vibronic spectra for Nile Red in explicit solvent environments\cite{Zuehlsdorff2018a}. Due to the high computational cost associated with the large number of TDDFT calculations, the solvent environment was not treated quantum mechanically in the calculations involving Nile red, but was represented by classical point charges taken directly from the force field parameters. This simplified treatment of the environment can be expected to have negligible influence on the computed spectral density for Nile red in benzene\cite{Zuehlsdorff2020}. For Nile red in water, it is expected that the classical treatment of the environment yields an underestimation of the spectral weight in the low-frequency region of the spectral density but does not alter the high frequency region\cite{Zuehlsdorff2020}.

\section{Calculating Optical Spectra Using a Truncated Cumulant Expansion}
\label{app:optical_spectra}
The linear absorption spectra reported in this work are expressed as the following Fourier transform,
\begin{equation}
    \label{eqn:linear_abs_spectrum}
    \sigma(\omega) = \alpha(\omega)\int_{-\infty}^\infty e^{\textrm{i}\omega t}\chi(t),
\end{equation}
where we set the $\omega$-dependent prefactor $\alpha(\omega)=1$. The linear response function $\chi(t)$ is expressed in terms of a cumulant expansion truncated at 2nd order\cite{Zuehlsdorff2019, Mukamel1995,Mukamel1985}:
\begin{equation}
    \label{eqn:linear_response}
    \chi(t)=|\mu_\textrm{eg}|^2 e^{\textrm{i}\omega_\textrm{eg}^\textrm{av}t-g_2(t)},
\end{equation}
where $\mu_\textrm{eg}$ is the transition dipole moment between the electronic ground and excited state, $\omega_\textrm{eg}^\textrm{av}$ is the thermal average of the vertical excitation energy between ground and excited state, and $g_2(t)$ is the 2nd order cumulant lineshape function. Note that the Condon approximation \cite{Condon1926, Condon1928} is applied, i.e. the transition dipole moment $\mu_\textrm{eg}$ is assumed to be independent of the nuclear positions. The lineshape function $g_2(t)$ can be expressed in terms of the spectral density $\mathscr{J}(\omega)$, such that \cite{Zuehlsdorff2019,Mukamel1995,Mukamel1985}
\begin{equation}
\begin{split}
    \label{eqn:lineshape}
    g_2(t)=\frac{1}{\pi}\int_0^\infty \textrm{d}\omega\, \frac{\mathscr{J}(\omega)}{\omega^2}\left[\textrm{coth}\left(\frac{\beta \omega}{2}\right)\left[1-\cos{\omega t}\right] -\textrm{i}\left[\sin{\omega t}-\omega t\right] \right].
    \end{split}
\end{equation}
Formally, the spectral density is a functional of the quantum autocorrelation function $C_{\delta U}(t)$ of the energy gap fluctuation operator $\delta U$, where $\delta U(\boldsymbol{\hat{q}})=H_e(\boldsymbol{\hat{q}})-H_g(\boldsymbol{\hat{q}})-\omega_\textrm{eg}^{\textrm{av}}$, and $H_g$ and $H_e$ are the nuclear Hamiltonians of the electronic ground- and excited-state potential energy surface respectively. Since the full quantum correlation function is in general inaccessible in atomistic condensed phase systems, we relate $\mathscr{J}(\omega)$ to the classical autorcorrelation function of energy gap fluctuations $C^\textrm{cl}_{\delta U}(t)$ that can be constructed from the vertical excitation energies computed along an MD trajectory\cite{Bader1994,Kim2002,Mukamel1995,Egorov1999}. This is achieved by using the harmonic quantum correction factor\cite{Craig2004,Ramirez2004} such that
\begin{equation}
    \label{eqn:spectral_density}
    \mathscr{J}(\omega)=\theta(\omega)\frac{\beta\omega}{2}\int_{-\infty}^\infty \textrm{d}t\,e^{\textrm{i}\omega t}C_{\delta U}^{\textrm{cl}}(t)e^{-|t|/\tau},
\end{equation}
where $\theta(\omega)$ is the Heaviside step function and the decaying exponential $e^{-|t|/\tau}$ is introduced to guarantee that the Fourier transform of $C^\textrm{cl}_{\delta U}(t)$ is well-behaved. For all calculations reported in this work, a decay constant of $\tau=500$~fs was used. 
The 2DES signals reported in this work are directly related to the third-order response function $\chi(t_3,t_2,t_1)$. The response function can be written as the sum of four terms, which, under the 2nd order cumulant approximation, can be expressed through the 2nd order lineshape function $g_2(t)$ \cite{Mukamel1995}:
\begin{equation}
    \label{eqn:third_order_response}
    \begin{split}
        R_1(t_3,t_2,t_1)=&\textrm{exp}[-g_2(t_1)-g_2^*(t_2)-g_2^*(t_3)+g_2(t_1+t_2)\\
        &+g^*_2(t_2+t_3)-g_2(t_1+t_2+t_3)],\\
        R_2(t_3,t_2,t_1)=&\textrm{exp}[-g^*_2(t_1)+g_2(t_2)-g_2^*(t_3)-g^*_2(t_1+t_2)\\
        &-g_2(t_2+t_3)+g^*_2(t_1+t_2+t_3)], \\
        R_3(t_3,t_2,t_1)=&\textrm{exp}[-g^*_2(t_1)-g_2(t_2)-g_2(t_3)-g^*_2(t_1+t_2)\\
        &-g^*_2(t_2+t_3)+g^*_2(t_1+t_2+t_3)], \\
        R_4(t_3,t_2,t_1)=&\textrm{exp}[-g_2(t_1)-g_2(t_2)-g_2(t_3)+g_2(t_1+t_2)\\
        &+g_2(t_2+t_3)-g_2(t_1+t_2+t_3)].
    \end{split}
\end{equation}
The 2DES spectra reported in this work are the purely absorptive spectra, obtained by adding the rephasing ($R_2$ and $R_3$) and non-rephasing ($R_1$ and $R_4$) contributions to the response function and Fourier-transforming the $t_1$ and $t_3$ time variables. For a given delay time $t_2=t_\textrm{delay}$, the absorptive 2DES spectrum can then be expressed as:
\begin{equation}
\label{eqn:2DES}
\begin{split}
S_{\textrm{2DES}}(\omega_3,t_\textrm{delay},\omega_1)\propto \textrm{Re} \int_0^\infty\textrm{d}t_3\int_0^\infty\textrm{d}t_1 \times\\
\big[e^{\textrm{i}\omega_3t_3+\textrm{i}\omega_1t_1}
\left(R_1(t_3,t_\textrm{delay},t_1)+R_4(t_3,t_\textrm{delay},t_1)\right) \\
 e^{-\textrm{i}\omega_3t_3+\textrm{i}\omega_1t_1}\left(R_2(t_3,t_\textrm{delay},t_1)+R_3(t_3,t_\textrm{delay},t_1)\right)\big].
\end{split}
\end{equation}

\section{Machine Learning Details}
\label{app:ml}
\begin{figure}[hbt!]
	\begin{center}
		\includegraphics[width=0.8\textwidth]{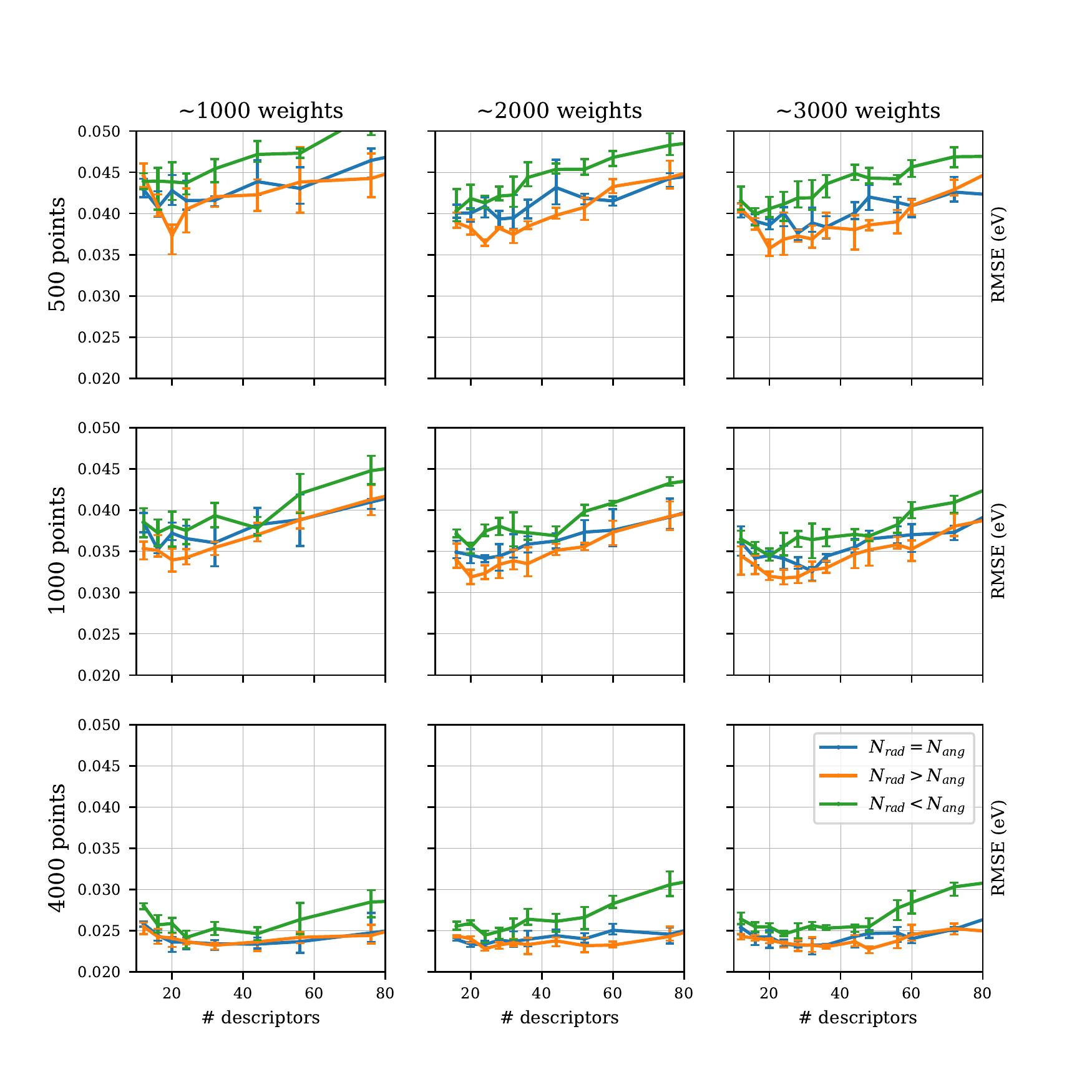}
	\end{center}
    \caption{Hyperparameter scans for $\text{pCT}^-$ in vacuum. Scan of average validation RMSE (the y-axes) as a function of the number of total weights in the model (columns) and amount of training data (rows). For a given number of total weights the number of descriptors is scanned (x-axes). The lines represent different ratios of radial $N_{rad}$ to angular $N_{ang}$ descriptors: 1:1 (blue), 2:1 (orange), 1:2 (green). The hyperparameters for the model with $\sim$3000 and 20 descriptors split 2:1 radial to angular were used for production.}
	\label{fig:pyp_vac_hyperparameter_scan}
\end{figure}

Given the number of hyperparameters involved in the specification of the ML models, and that they are all interdependent to some degree, we chose to scan only a subset of them and keep others fixed. For instance, the cutoffs for the radial and angular Chebyshev polynomial descriptors were set to 5~\AA\, for all of the models we used. For $\text{pCT}^-$ in water there are on average 200 solvent atoms within 5~$\AA$ of any chromophore atom (186 for Nile red in benzene and 221 for Nile red in water). On the other hand, we scanned the hyperparameters that directly specified the number of total weights in our models. The total number of weights is given by the specification of how many input nodes, hidden layers, and hidden nodes per hidden layer are in our neural networks. We used the same number of hidden nodes for each of the two hidden layers. The number of input nodes (i.e. number of descriptors) was determined by the choice of radial and angular expansion orders for the Chebyshev polynomial descriptors.

In each of our hyperparameter scans, as depicted in SI Figure~\ref{fig:pyp_vac_hyperparameter_scan}, we scanned the number of input nodes while keeping the total number of weights roughly the same. This meant that for a given number of total weights, we increased the number of input nodes while decreasing the number of hidden nodes accordingly. Also, in comparing the different columns, the increase in total number of weights for a given number of descriptors is a result of increasing the number of hidden nodes accordingly. Note that a particular number of input nodes can be specified using different ratios of radial to angular expansion orders (e.g. 2:1, 1:1, 1:2). We performed these scans for models with $\sim$1000, $\sim$2000, and $\sim$3000 total weights as fitted to 500, 1000, and 4000 energy gaps for $\text{pCT}^-$ in vacuum. Note that the total weights for a given mode (e.g. $\sim$2000) is a sum of the number of weights in each of the four element-specific neural networks for $\text{pCT}^-$ in vacuum (e.g. $\sim$500). From the hyperparameter scans, we see that a 2:1 radial to angular expansion order systematically outperforms the other two ratios. We also found that when trained on 500 points, the model with the lowest average validation RMSE was the one with $\sim$3000 total weights and 20 input nodes (2:1 radial to angular) and that this choice of hyperparameters performed well when also trained on larger datasets. Hence, we settled on this specific choice of hyperparameters for $\text{pCT}^-$ in vacuum, which corresponds to a radial expansion order of 6, an angular expansion order of 2, and 18 hidden nodes per hidden layer for each of the element-specific neural networks. This model has a total of 2956 fitting parameters. Note that the average validation RMSEs reported in SI Figure~\ref{fig:pyp_vac_hyperparameter_scan}, and in all other learning curves we present, are calculated as a mean over four different randomly initialized models and are evaluated using a fixed validation set of 4000 points. The $\text{tanh}$ activation function was used for each of the hidden nodes and a linear activation function was used for the output node. A modified version of the {\ae}net software package \cite{Artrith2016} was used for all of the energy gap fitting and prediction. To demonstrate that our choice of hyperparameters is transferable across systems we used the same set for all of the systems we examined.
\begin{figure}[hbt!]
	\begin{center}
		\includegraphics[width=0.7\textwidth]{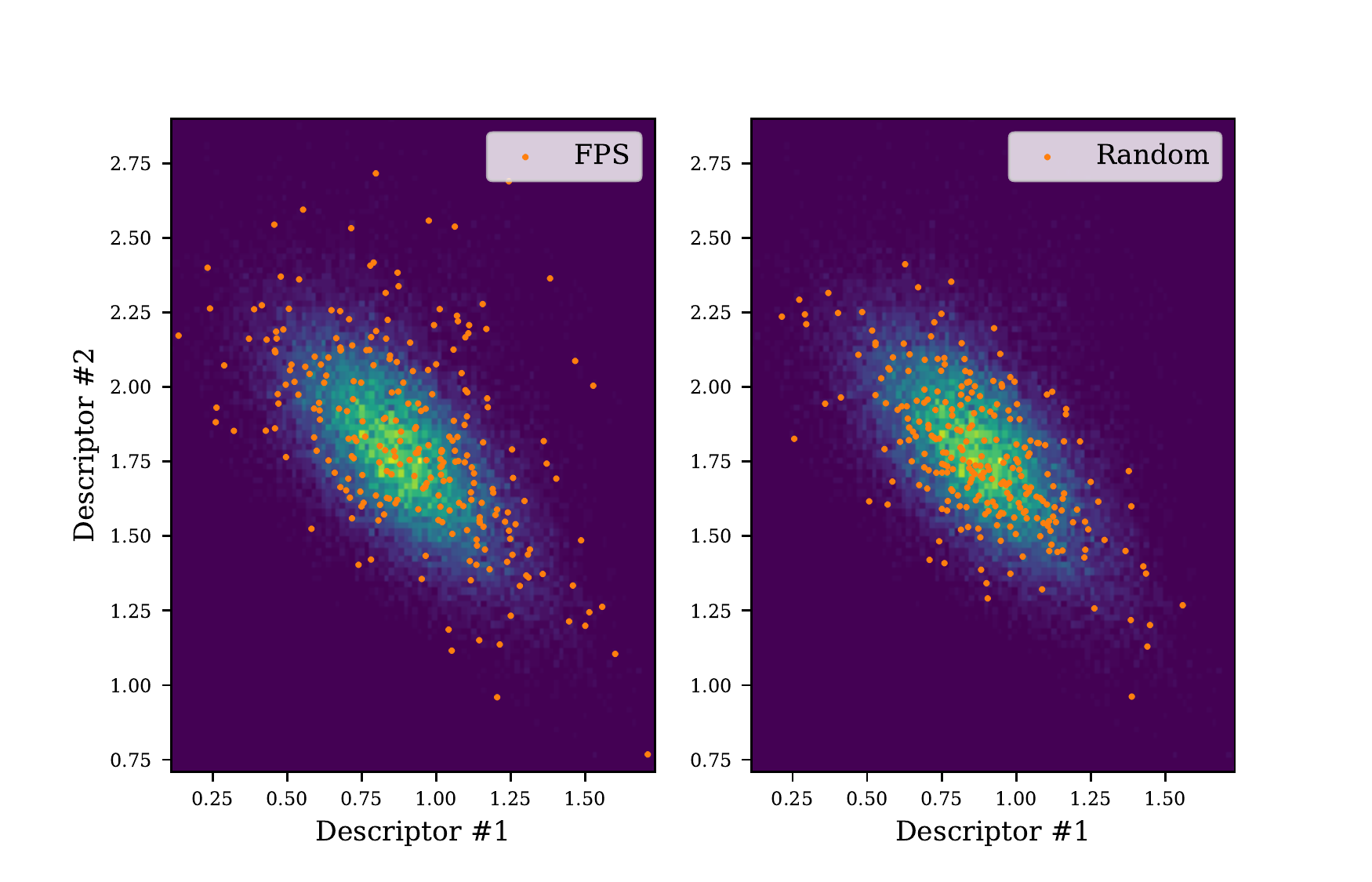}
	\end{center}
    \caption{Comparison of farthest-point sampling (left) vs. random sampling (right) for $\text{pCT}^-$ in vacuum. For visualization purposes, the data points are projected onto the space spanned by the first and second chromophore descriptors. The 250 points selected from the overall dataset of 36,000 by farthest-point sampling are more diverse. }
	\label{fig:pyp_vacuum_fps_vs_random}
\end{figure}

For each system, 4000 randomly selected points were withheld from the fitting procedure and reserved as a validation set. Training points were selected amongst the remaining points using farthest-point sampling\cite{Imbalzano2018}, as conducted in the space spanned by the chromophore descriptors. More specifically, we define a chromophore descriptor set to be the ordered concatenation of all chromophore atom descriptors. Note that the ordering is arbitrary but consistent for a given system. For instance, the $\text{pCT}^-$ chromophore consists of 29 atoms each of which has 20 associated atomic Chebyshev polynomial descriptors, so the concatenated chromophore descriptor will consist of 580 values. We ran farthest-point sampling in this space in order to select a diverse set of training points from the overall dataset. For $\text{pCT}^-$ in vacuum, a representative sampling of 500 points from the full set of 36,000 possible training points is shown in SI Figure~\ref{fig:pyp_vacuum_fps_vs_random}. The points have been projected onto the subspace spanned by the first and second chromophore descriptors for this system. In short, farthest-point sampling selects a wider distribution of training points than random selection. Note that the individual atomic descriptors for a given element type were normalized to zero mean and unit standard deviation, as is standard in aenet. When these are combined into a chromophore descriptor, no further normalization is applied. Thus the means are no longer necessarily zero and the standard deviations are no longer necessarily one, as we see in SI Figure~\ref{fig:pyp_vacuum_fps_vs_random}.
\begin{figure}[hbt!]
	\begin{center}
		\includegraphics[width=0.4\textwidth]{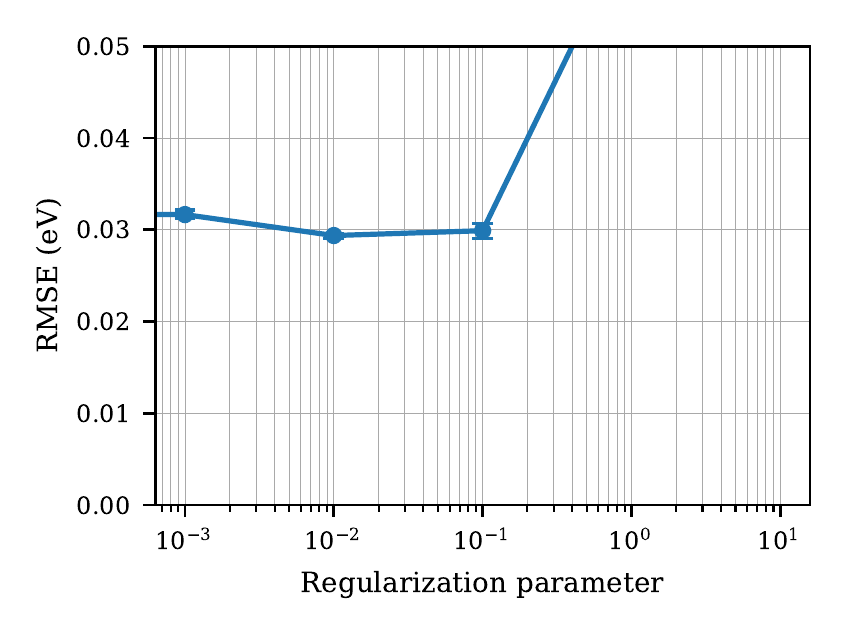}\includegraphics[width=0.4\textwidth]{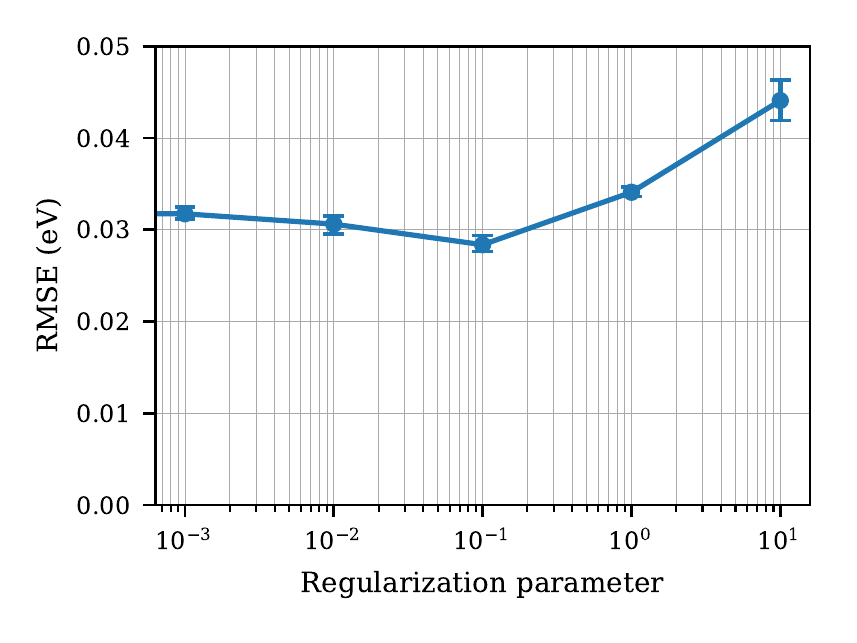}
	\end{center}
    \caption{Average validation RMSEs when using L1 and L2 regularization for models trained on 1000 points for $\text{pCT}^-$ in vacuum. Neither method provides much in terms of accuracy benefits.}
	\label{fig:pyp_vac_regularization}
\end{figure}

Optimization of the fitting parameters was conducted using L-BFGS \cite{Byrd1995} with no explicit regularization of the cost function. Early stopping was employed in order to avoid overfitting. Other explicit regularization techniques such as L1 and L2 regularization were also tested for $\text{pCT}^-$ in vacuum and the corresponding average validation RMSEs when trained on 1000 data points are reported in SI Figure~\ref{fig:pyp_vac_regularization}. Explicit regularization only marginally improved accuracies over the early stopping result, from a test RMSE of 0.0318 eV without regularization to 0.0284 with L2 using a regularization parameter of 0.1, i.e. a improvement of $\sim$10\% that leads to minimal changes in the computed spectra. We decided that this slight improvement did not warrant their usage since it would involve one more tunable hyperparameter.

\section{Reorganization energies}
\label{app:reorg_energy}
\begin{table}[htb!]
    \centering
    \begin{tabular}{c|c|c|c|}
             & Direct-solvent & Indirect-solvent & Hidden-solvent \\ \hline
        8000 & 0.277 & 0.278 & 0.253 \\ \hline
        2000 & 0.260 & 0.262 & 0.232 \\ \hline
        500  & 0.190 & 0.194 & 0.236
    \end{tabular}
    \caption{Reorganization energies in eV for $\text{pCT}^-$ in water as calculated using the energy gap predictions from the three ML solvation models trained on 8000, 2000, and 500 points. For comparison, the reference reorganization energy calculated using TDDFT energy gaps was $\lambda = 0.284 \text{eV}$ whereas the reorganization energy from TDDFT energy gaps for $\text{pCT}^-$ in vacuum was calculated to be $\lambda=0.089 \text{eV}$.}
    \label{tab:pyp_reorganization_energies}
\end{table}

The dynamic Stokes shift observed in the 2DES of $\text{pCT}^-$ was primarily attributed to reorganization of the system's degrees of freedom, both intrachromophore and solvent-coupled, to stabilize the excited state electronic density. This stabilization energy is referred to as a reorganization energy $\lambda$ and is given by \cite{Mukamel1995},
\begin{equation}
    \label{eqn:reorganization_energy}
    \lambda=\frac{1}{\pi}\int_{0}^\infty \textrm{d}\omega \frac{\mathscr{J}(\omega)}{\omega}.
\end{equation}
Generally, one expects the reorganization energy to be greater for systems where there is a stronger interaction between chromophore and solvent.

\section{$\text{pCT}^-$ in Vacuum}
\label{app:pyp_vac}
\begin{figure}[hbt!]
	\begin{center}
	    \includegraphics[width=0.5\textwidth]{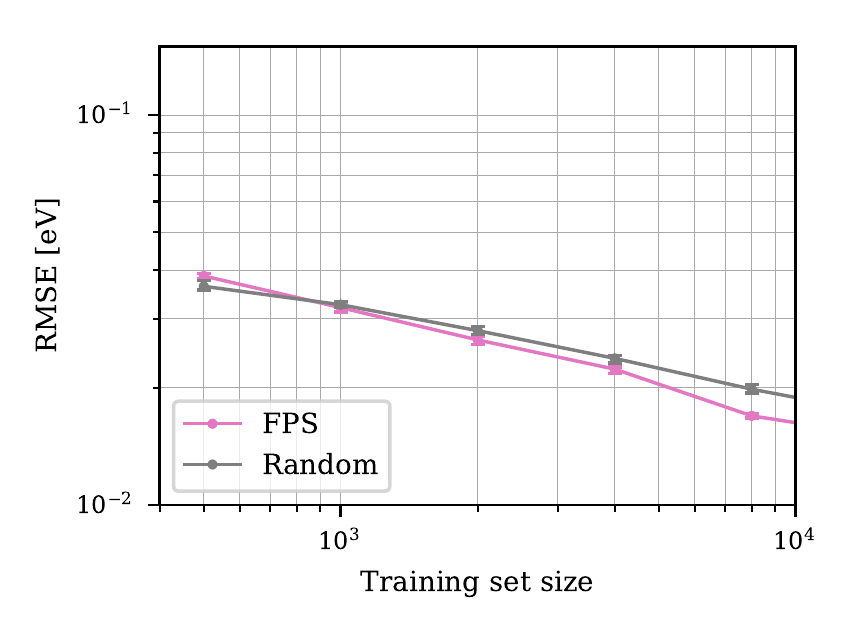}
		\includegraphics[width=0.5\textwidth]{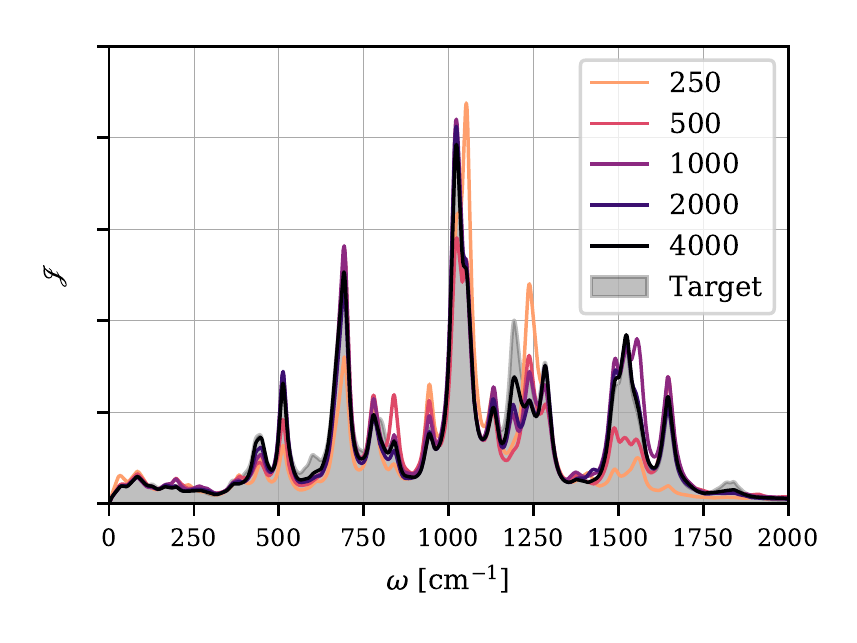}\includegraphics[width=0.5\textwidth]{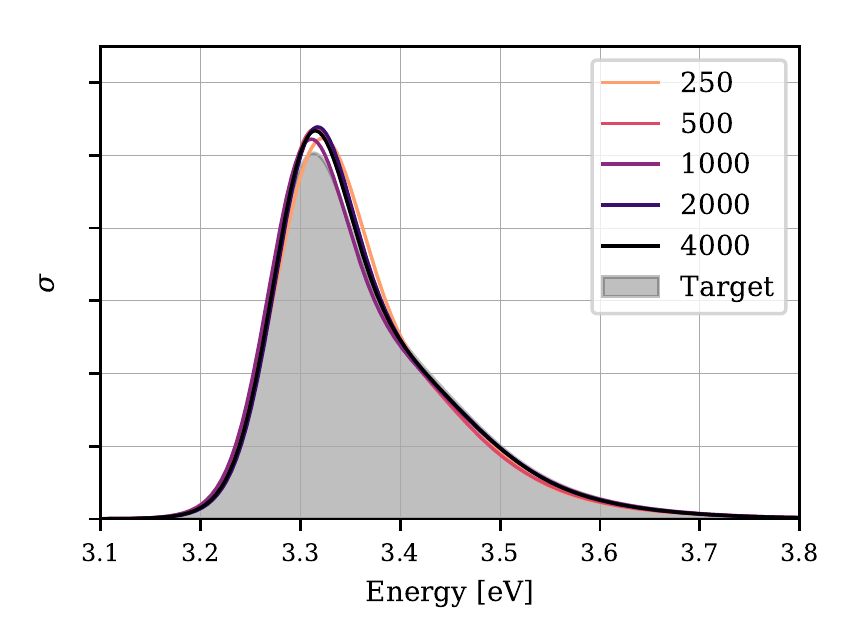}
	\end{center}
    \caption{Learning curve (top) of average validation RMSEs, spectral densities (bottom left), and linear absorption spectra (bottom right) for $\text{pCT}^-$ in vacuum as a function of training set sizes. Spectral densities and linear absorption spectra were constructed using 40,000 energy gaps. The TDDFT results serve as the reference targets. The linear absorption spectrum is qualitatively accurate for as few as 500 training points.}
	\label{fig:pyp_vac}
\end{figure}
SI Figure~\ref{fig:pyp_vac} shows the learning curves for models trained on datasets selected using farthest-point sampling and random sampling for $\text{pCT}^-$ in vacuum. The average validation RMSEs are comparable and farthest-point sampling does not provide significant benefits here. This hints at the possibility that certain high-variance descriptors do not correlate much with the chromophore's energy gap. Still, we chose to use farthest-point sampling for all of our results since the corresponding models, when trained on 1000 or more points, were slightly more accurate and using farthest-point sampling did not require any additional hyperparameters. The corresponding spectral densities and linear absorption spectra, when fitted on farthest-point sampling selected training sets, are also shown in SI Figure~\ref{fig:pyp_vac}. With as few as 500 points, we are able to train an ML model that accurately captures the linear absorption spectrum for this system.

\section{Correlation plots for $\text{pCT}^-$ in water}
\label{app:pyp_water_corrplots}
\begin{figure}[hbt!]
	\begin{center}
		\includegraphics[width=0.75\textwidth]{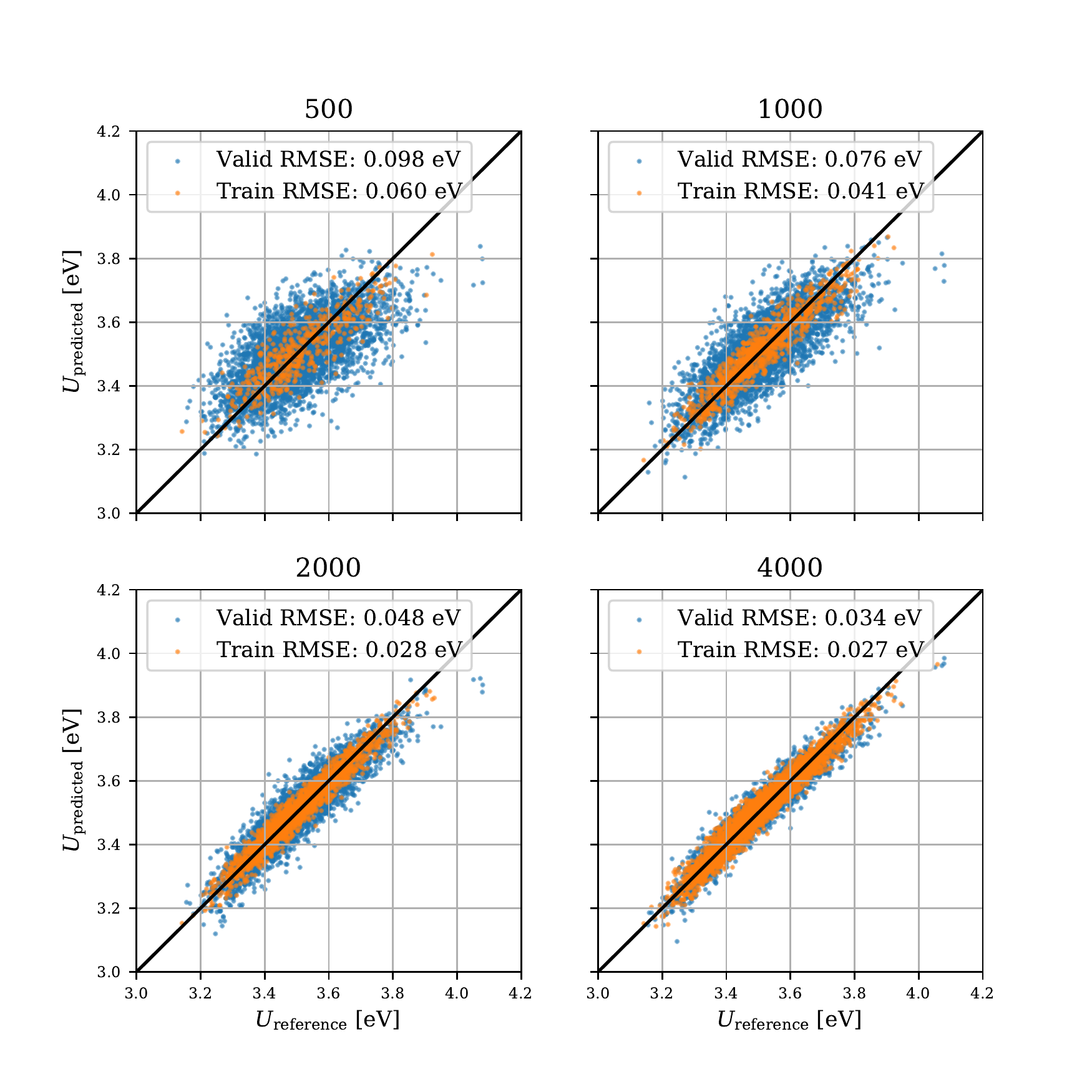}
	\end{center}
    \caption{Plots depicting how the indirect-solvent predicted energy gaps $U_\mathrm{predicted}$ and the reference TDDFT energy gaps $U_\mathrm{reference}$ become increasingly correlated as the training set size is increase from 500 to 4000 points. The same validation set was used for each plot and consisted of 4000 points.}
	\label{fig:pyp_water_corrplots}
\end{figure}

The correlation plots in SI Figure~\ref{fig:pyp_water_corrplots} provide a more detailed point-by-point look at how the ML predicted energy gaps compare with those of the reference. Only the predictions from the fit with the smallest validation RMSE for each training set size are shown. Note that the projection and histogramming of these values onto their respective y-axis (x-axis) would give us the ensemble absorption spectrum based on the ML predictions (reference calculations).

To provide additional context for the RMSEs that are reported, note that the reference energy gaps for this system range from 3.101-4.207~eV and the standard deviation is 0.123~eV.

\section{2DES for $\text{pCT}^-$ in water as a function of training set size}
\label{app:pyp_water_2DES}
\begin{figure}[hbt!]
	\begin{center}
		\includegraphics[width=0.5\textwidth]{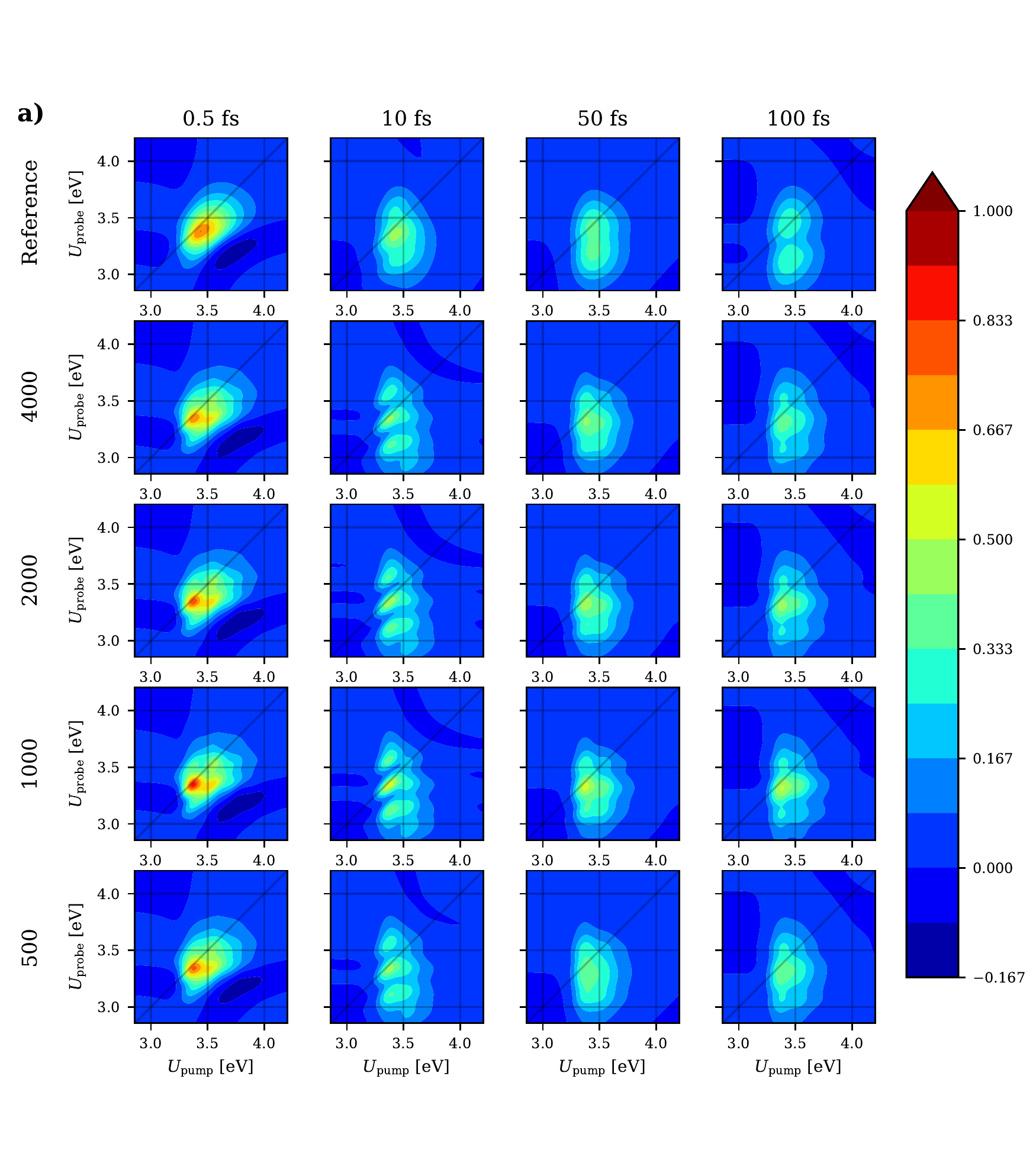}\includegraphics[width=0.5\textwidth]{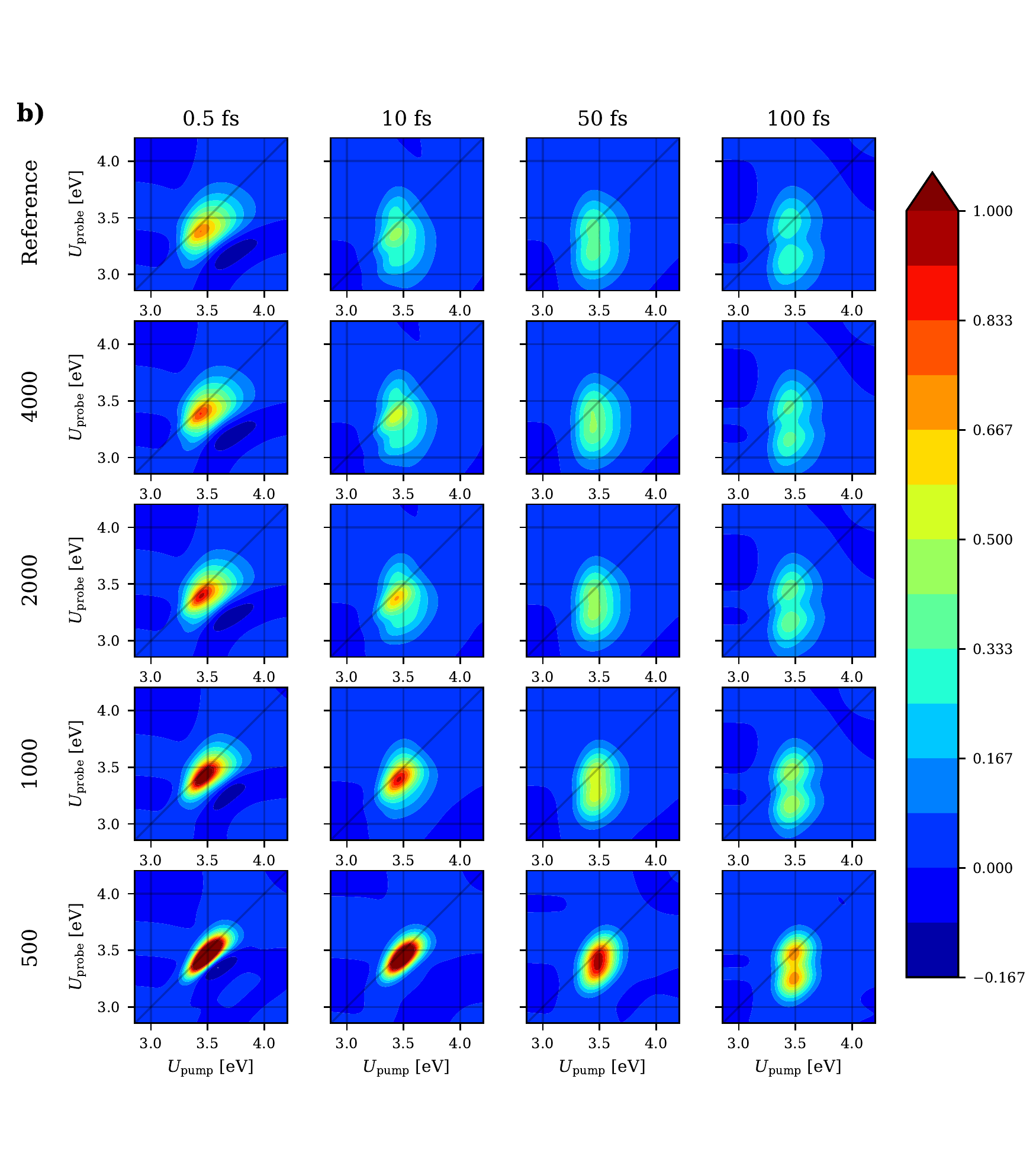}
	\end{center}
    \caption{2DES of $\text{pCT}^-$ in water for hidden-solvent models (a) and indirect-solvent models (b) trained on 4000, 2000, 1000, 500 points (rows) as compared to the reference 2DES (top row) for time delays of 0.5, 10, 50, 100~fs (columns). Spectra were computed using 32,000 energy gaps with the TDDFT results as reference.}
	\label{fig:pyp_water_2DES_trainingsetsize}
\end{figure}

From SI Figure~\ref{fig:pyp_water_2DES_trainingsetsize}b, we see for the indirect-solvent models that there is a decay in the accuracy of the 2DES for $\text{pCT}^-$ in water as the training set size is reduced from 4000 to 500 points. This is most prominent for the time delay of 10~fs where we still see off-diagonal vibronic couplings in the reference but these are lost as we reduce the training size. This is consistent with what we saw in the corresponding spectral densities in Figure~4 of the main text where the intensities of the high-frequency intrachromophore modes were heavily underestimated by the indirect-solvent models trained on only 500 and 1000 points. Interestingly, and in accord with what we showed in the main text for the linear absorption spectra (Fig.~3), the performance of the hidden-solvent model (SI Fig.~\ref{fig:pyp_water_2DES_trainingsetsize}a) did not see as much degradation going from 4000 to 500 training points. 

\section{2DES for Nile red in benzene as a function of training set size}
\label{app:nilered_benzene_2DES}
\begin{figure}[hbt!]
	\begin{center}
		\includegraphics[width=0.5\textwidth]{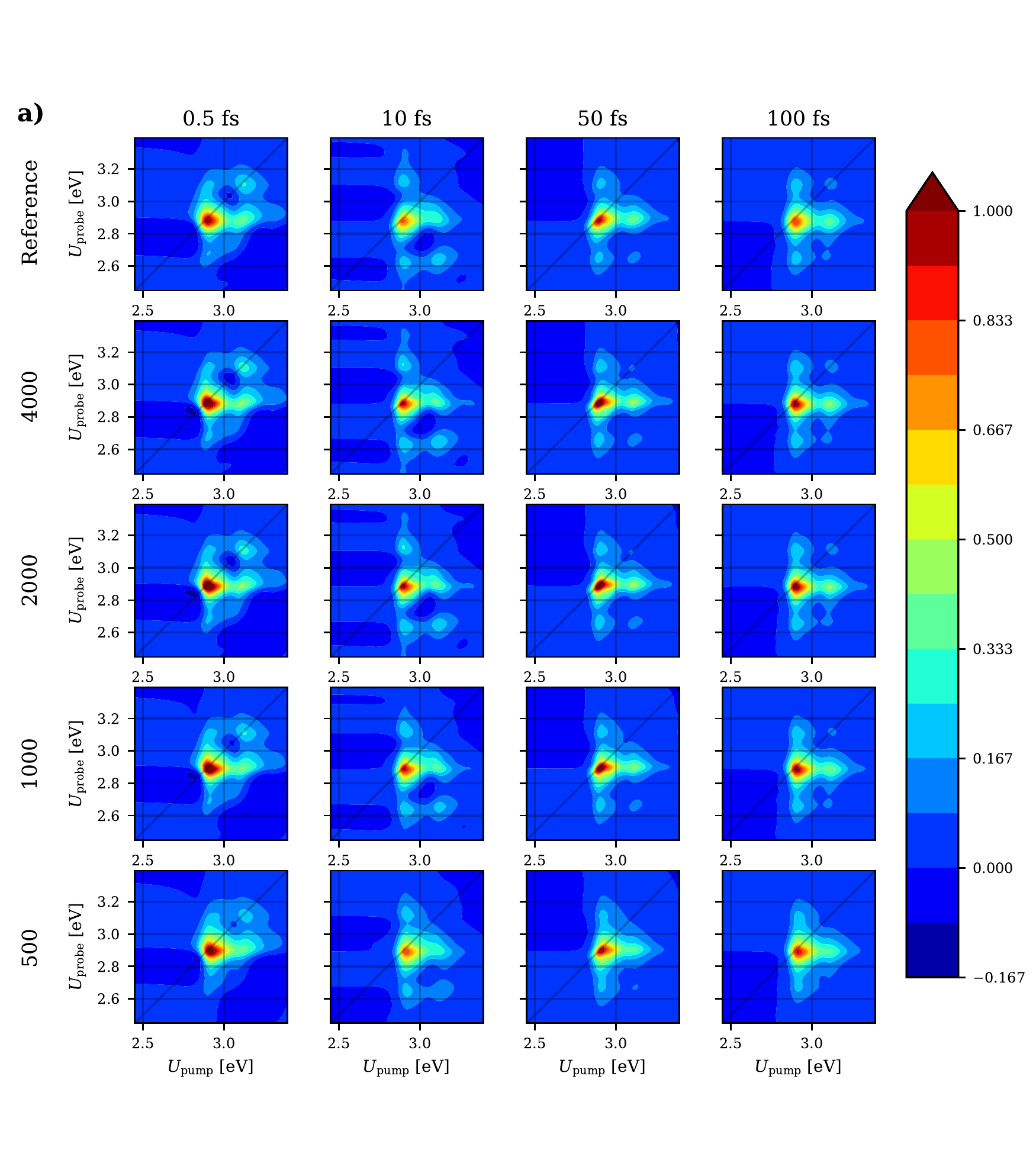}\includegraphics[width=0.5\textwidth]{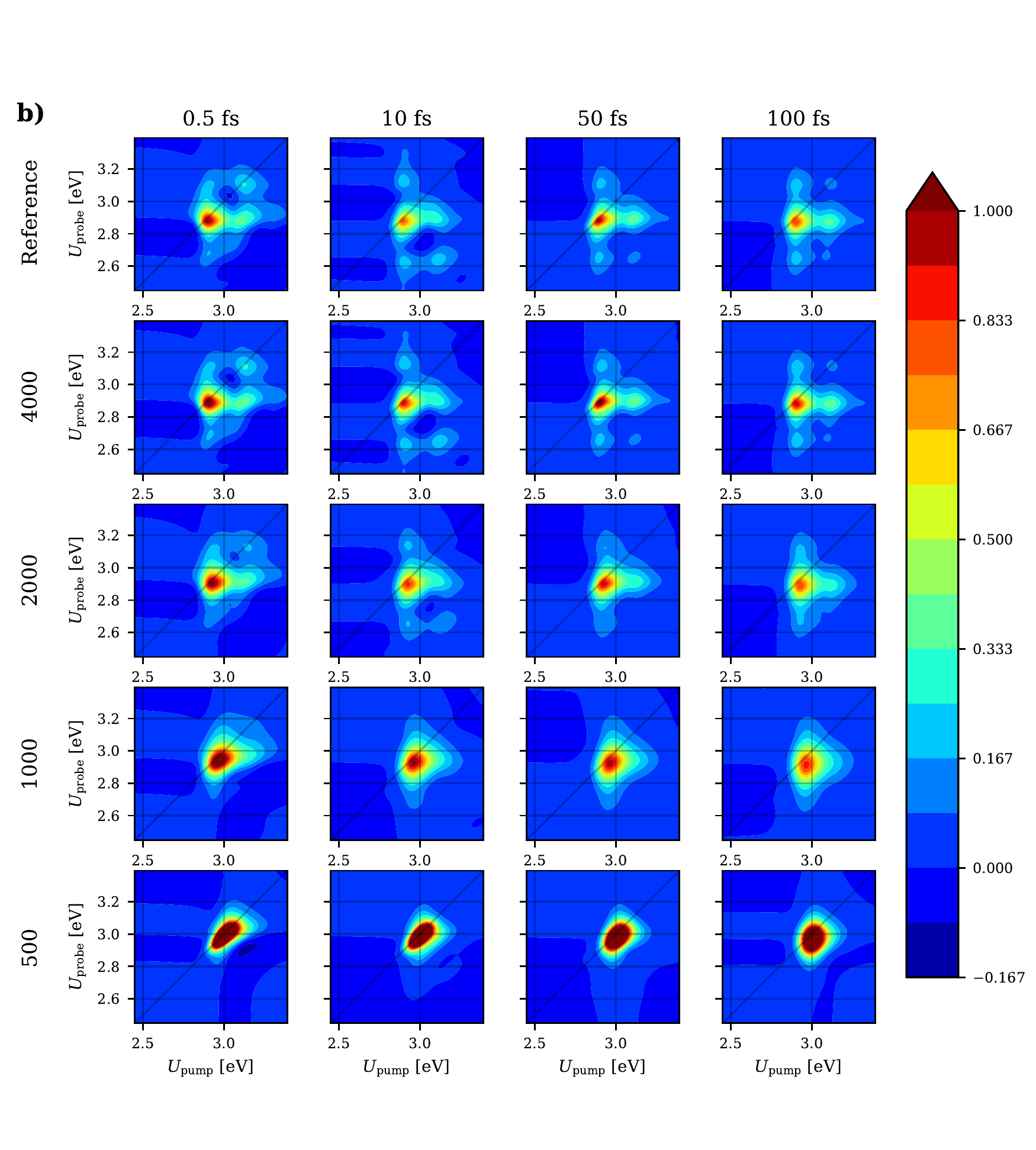}
	\end{center}
    \caption{2DES of Nile red in benzene for hidden-solvent models (a) and indirect-solvent models (b) trained on 4000, 2000, 1000, 500 points (rows) as compared to the reference 2DES (top row) for time delays of 0.5, 10, 50, 100~fs (columns). Spectra were computed using 30,000 energy gaps with the references being the TDDFT results.}
	\label{fig:nilered_benzene_2DES_trainingsetsize}
\end{figure}

The 2DES for Nile red in benzene (SI Fig.~6) has peaks characteristic of a 4-level vibronic system where the ground $S_0$ and first electronic excited state $S_1$ each have two accessible vibrational levels denoted as $\nu_0$ and $\nu_1$\cite{Provazza2017}. There are roughly three vibronic energy scales in this system, representing from highest to lowest the transitions $S_1^{\nu_1} - S_0^{\nu_0}$, $S_1^{\nu_1}-S_0^{\nu_1}$ or $S_1^{\nu_0} - S_0^{\nu_0}$, and $S_1^{\nu_0}-S_0^{\nu_1}$. The transitions corresponding to those energy scales are what give rise to the gridded pattern for the 2DES.

From SI Figure~\ref{fig:nilered_benzene_2DES_trainingsetsize}a, we see that the hidden-solvent model is able to capture the 2DES for Nile red in benzene with as few as 1000 training points. Even the result for the model trained on only 500 points looks qualitatively good aside from the blurred vibronic features. On the other hand, the accuracy of the 2DES computed using the indirect-solvent model degrades considerably going from 4000 to 500 training points (SI Fig.~\ref{fig:nilered_benzene_2DES_trainingsetsize}). The off-diagonal vibronic peaks become more blurred as training points are decreased, which results from underpredicting the high-frequency intrachromophore modes of the spectral density (Fig.~6). Qualitatively, the 2000 training point result using the indirect-solvent model is comparable to the 500 training point result using the hidden-solvent model.

\section{Linear absorption spectrum and 2DES for Nile red in water as a function of training set size}
\label{app:nilered_benzene_2DES}
\begin{figure}[hbt!]
	\begin{center}
		\includegraphics[width=0.5\textwidth]{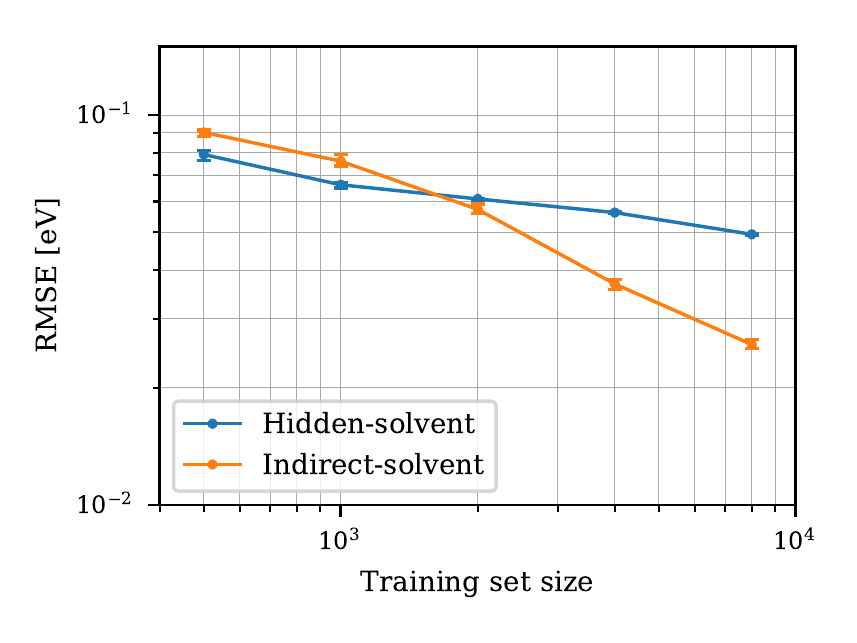}\includegraphics[width=0.5\textwidth]{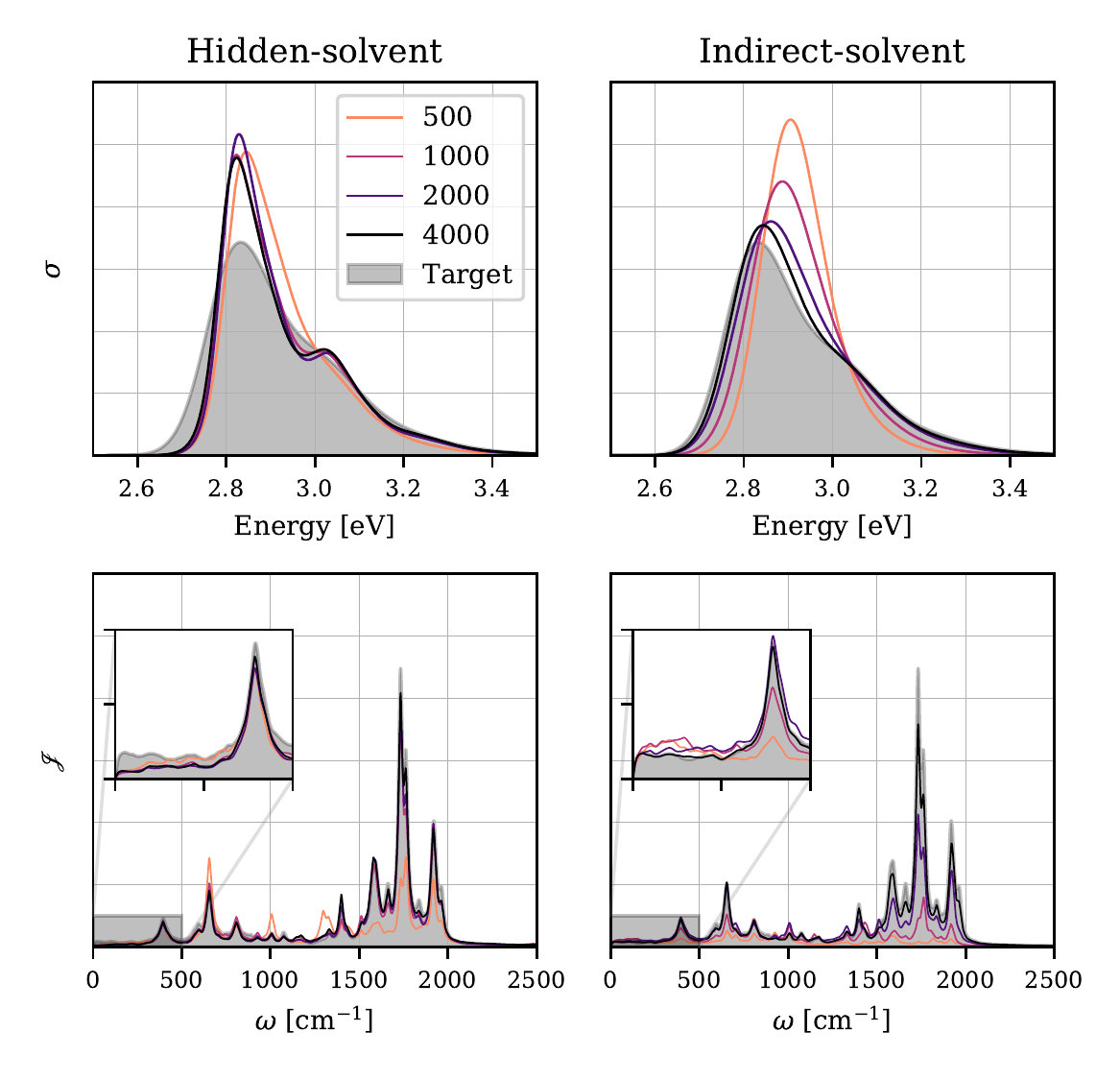}
	\end{center}
    \caption{Learning curves (left) of average validation RMSEs, linear absorption spectra (right, top row), and spectral densities (right, bottom row) of the hidden- and indirect-solvent models when trained on various training set sizes for Nile red in water. Linear absorption spectra and spectral densities were computed using 30,000 energy gaps with the TDDFT results used as reference.}
	\label{fig:nilered_water_lc_specdens_absspec}
\end{figure}

SI Figure~\ref{fig:nilered_water_lc_specdens_absspec} for Nile Red in water depicts trends similar to what we saw with $\text{pCT}^-$ in water. The accuracy of the hidden-solvent model, in neglecting solvent atom positions, is limited and hence results in linear absorption spectra where vibronic features are over-accentuated even when trained on 4000 training points. We see from the corresponding spectral densities, specifically the underprediction of low frequency intensities, that this is primarily due to an inability to accurately predict the contributions of  chromophore-solvent interactions to the energy gap. However, the indirect-solvent model is able to qualitatively capture these same chromophore-solvent interactions that the hidden-solvent model misses when trained on a reasonable number of training points ($\sim$4000).

\begin{figure}[hbt!]
	\begin{center}
		\includegraphics[width=0.5\textwidth]{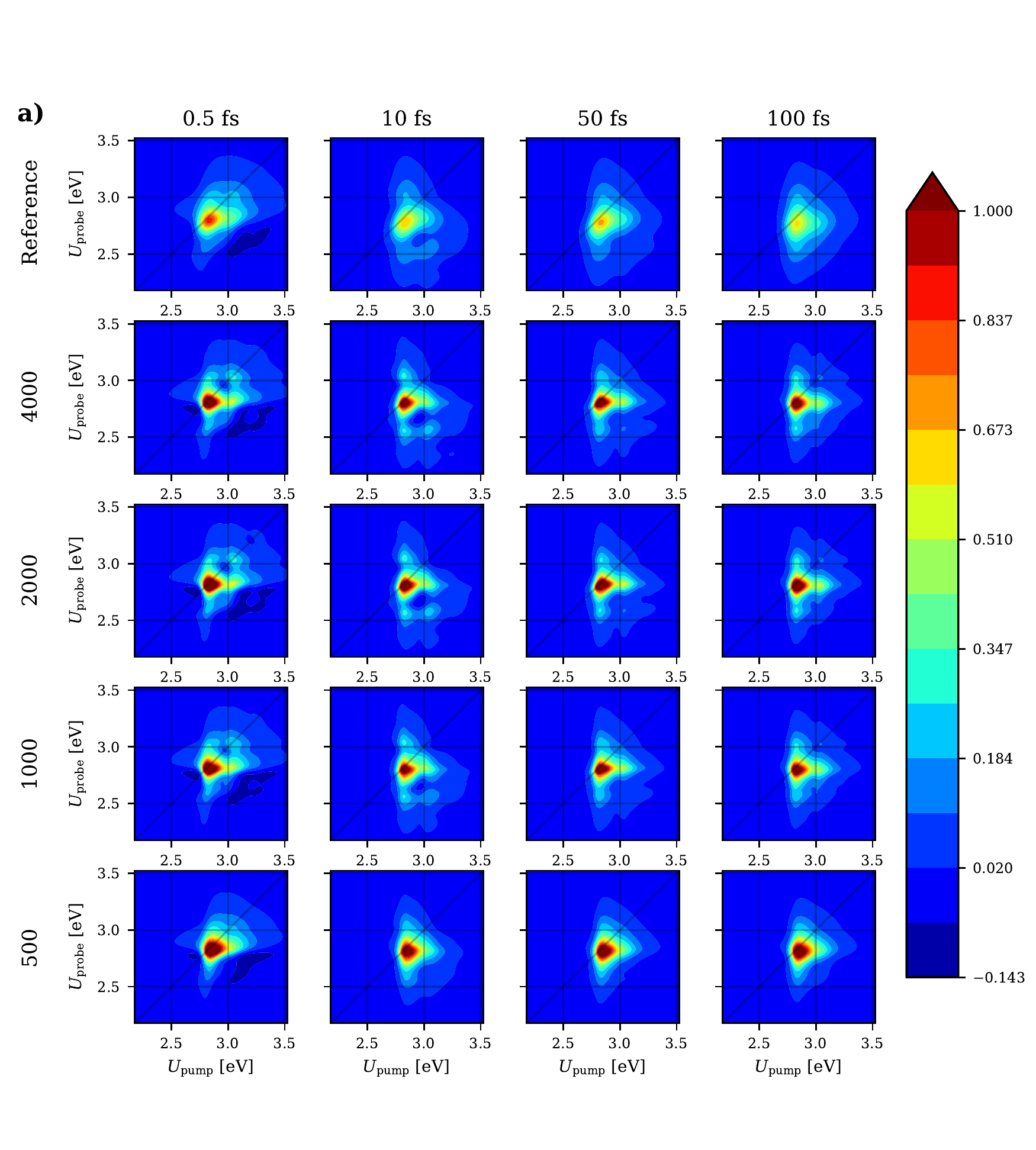}\includegraphics[width=0.5\textwidth]{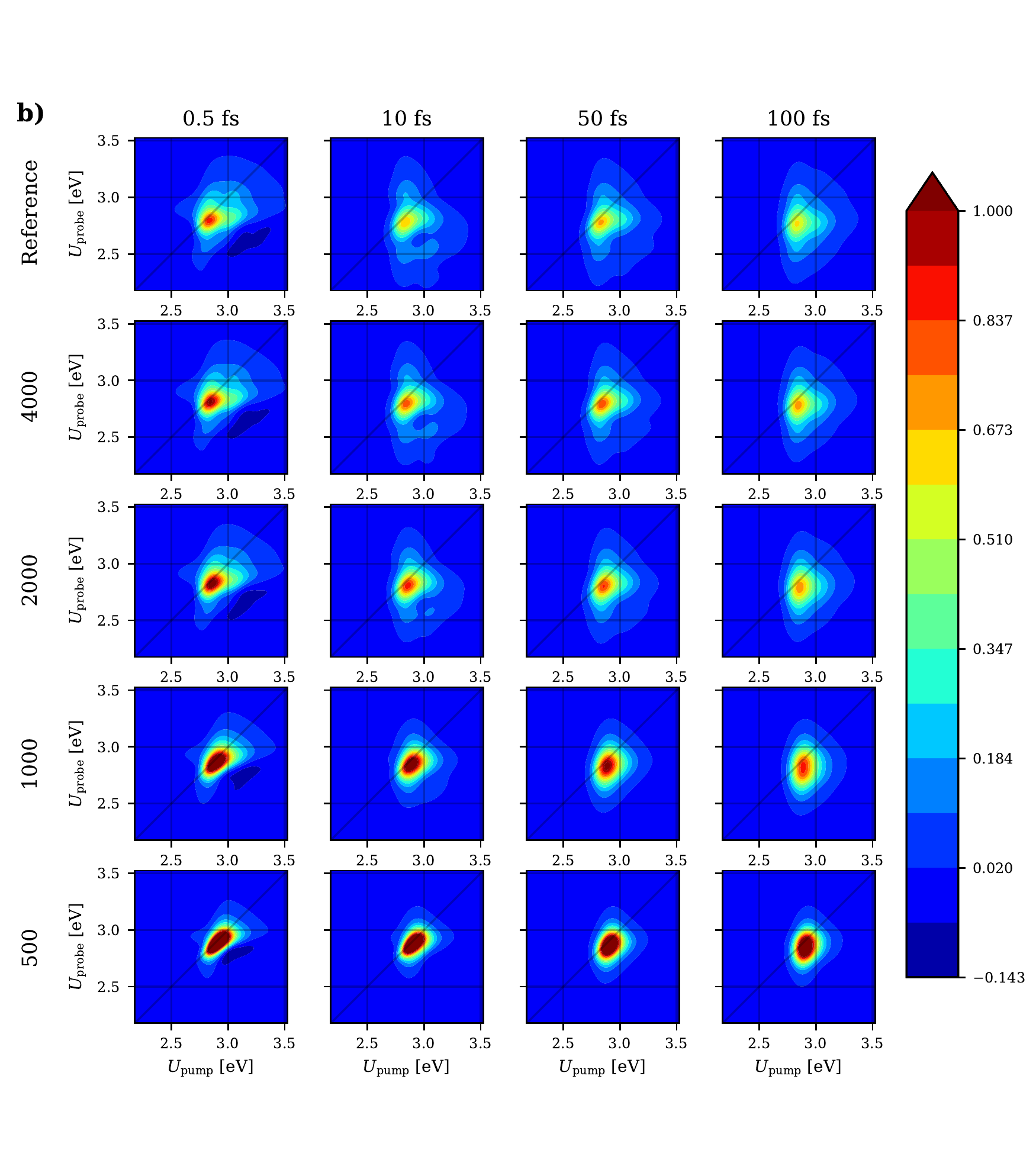}
	\end{center}
    \caption{2DES of Nile red in water for hidden-solvent models (a) and indirect-solvent models (b) trained on 4000, 2000, 1000, 500 points (rows) as compared to the reference 2DES (top row) for time delays of 0.5, 10, 50, 100~fs (columns). Spectra were computed using 30,000 energy gaps with the references being the TDDFT results.}
	\label{fig:nilered_water_2DES_trainingsetsize}
\end{figure}

The inaccuracy of the hidden-solvent model and the accuracy of the indirect-solvent model are similarly highlighted in the 2DES for Nile red in water (SI Fig.~\ref{fig:nilered_water_2DES_trainingsetsize}). Even with 4000 training points, the hidden-solvent model predictions result in a 2DES where vibronic features are spuriously resolvable due to an inability to correctly predict for broadening due to chromophore-solvent interactions. In fact, these 2DES resemble those of Nile red in benzene (SI Fig.~\ref{fig:nilered_benzene_2DES_trainingsetsize}) where the Nile red chromophore weakly interacts with the surrounding benzene molecules. Conversely, the indirect-solvent model is able to capture the 2DES for all time delays of Nile red in water when trained on 4000 points. However, the quality of the indirect-solvent 2DES degrades considerably when decreasing training set sizes. This is particularly evident in the loss of the off-diagonal vibronic peaks at 10~fs.

\section{Decomposition of reference 2DES for $\text{pCT}^-$ in water}
\label{app:pyp_water_2DES_pathways}
\begin{figure}[hbt!]
	\begin{center}
		\includegraphics[width=0.75\textwidth]{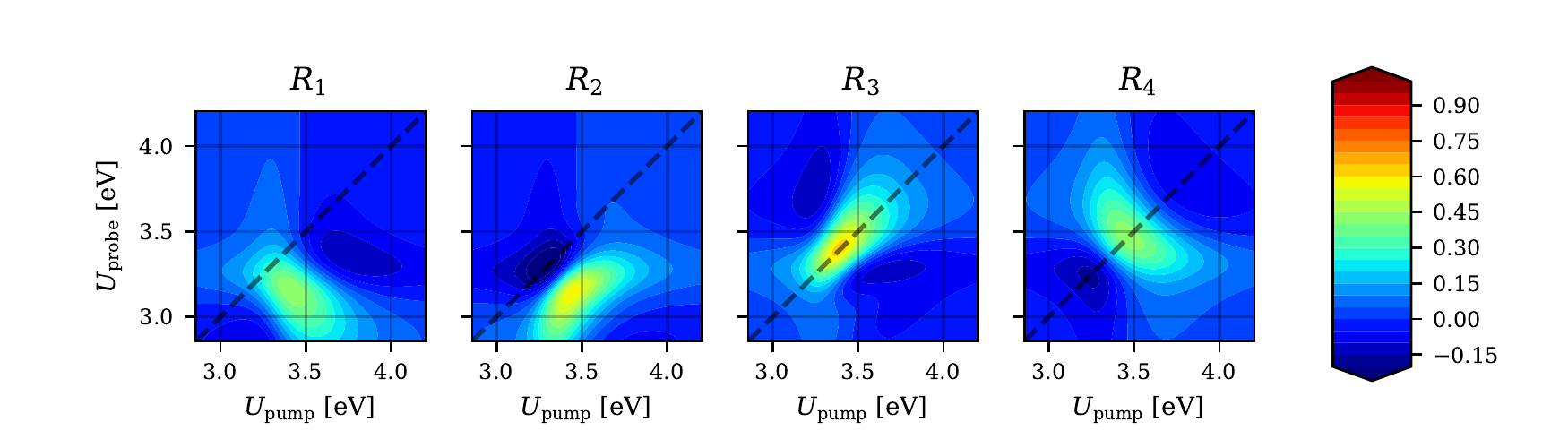}
	\end{center}
    \caption{The reference 2DES for $\text{pCT}^-$ in water at a time delay of 100~fs is decomposed into its four response function contributions. }
	\label{fig:pyp_water_2DES_pathways}
\end{figure}

The four individual contributions to the total third order response function (\ref{eqn:third_order_response}), denoted $R_1$, $R_2$, $R_3$, and $R_4$ (\ref{eqn:third_order_response}), correspond to different Liouville pathways. For a two-level system, these pathways can be unambiguously assigned to stimulated emission ($R_1$ and $R_2$) and ground-state bleaching ($R_3$ and $R_4$) processes\cite{Mukamel1995,Kwac2003}. By plotting these different contributions separately for $\text{pCT}^-$ in water when using a 100~fs 2DES time delay, as is done SI Figure~\ref{fig:pyp_water_2DES_pathways}, we see that $R_3$ and $R_4$ primarily contribute to the intensity of the diagonal peak and that $R_1$ and $R_2$ primarily contribute to the peak below the diagonal.